# A systematic literature review on the code smells datasets and validation mechanisms

Morteza Zakeri-Nasrabadi
School of Computer Engineering, Iran University of Science and Technology, Tehran, Iran, morteza_zakeri@comp.iust.ac.ir

Saeed Parsa[*]
School of Computer Engineering, Iran University of Science and Technology, Tehran, Iran, parsa@iust.ac.ir

Ehsan Esmaili
School of Computer Engineering, Iran University of Science and Technology, Tehran, Iran, eh_esmaili98@comp.iust.ac.ir

Fabio Palomba
Department of Computer Science, University of Salerno, Salerno, Italy, fpalomba@unisa.it

The accuracy reported for code smell-detecting tools varies depending on the dataset used to evaluate the tools. Our survey of 45 existing datasets reveals that the adequacy of a dataset for detecting smells highly depends on relevant properties such as the size, severity level, project types, number of each type of smell, number of smells, and the ratio of smelly to non-smelly samples in the dataset. Most existing datasets support God Class, Long Method, and Feature Envy while six smells in Fowler and Beck's catalog are not supported by any datasets. We conclude that existing datasets suffer from imbalanced samples, lack of supporting severity level, and restriction to Java language.

## 1    INTRODUCTION

Code smells postulate the refactoring opportunities in the source code of software systems during the development and maintenance phases [1]. Therefore, decent refactoring mainly depends on identifying existing smells in the target source code accurately and correctly. On the other hand, automatic smell detection requires a large dataset with annotated samples for each smell since most smell detection methods are based on statistical and learning-based approaches [2]. The performance of a code smell detection tool is vastly affected by the dataset used to create and evaluate that tool [3]. It has been shown that the results of some code smell detection tools are not accurate due to the method used in constructing and exploiting their datasets [3], [4]. For this reason, it is essential to identify the state-of-the-art datasets used for code smell detection, their capabilities, and their limitations.

A code smell dataset has two potential applications. First, it can be used as a data source to construct automatic software smell detection tools [5]. Second, it can be used as a ground truth or golden reference to evaluate a code smell detection tool [6]. The reliability of the dataset affects the performance of both applications. Identifying code smells is inherently a subjective process [7], requiring human experts' intervention. Code smell detection in industrial environments and tool evaluation in academic research heavily depend on human factors [8]. Without any standard and quality dataset, it is difficult to create code smell detection tools that are fully automated, and their results are not affected by human judgments.

Several studies have systematically reviewed the software smells, concepts, and detection tools [2], [9]–[12], while they have rarely focused on the details of the datasets and the validation techniques employed by researchers and practitioners. Di Nucci et al. [3] initially discussed the impact of the dataset on smell detection results with machine learning. They found out that the high performance reported in Fontana's work [13] is mainly related to the specific dataset employed rather than to the capabilities of ML techniques for code smell detection. It confirms the essence of the quality dataset in software smell detection [13]. However, they have not proposed any alternative dataset to mitigate these problems. Azeem et al. [2] state that only a few studies have used a large dataset to build and evaluate code smell detection tools. They do not discuss the type of smells and structures used in various datasets.

We observed that the code smell datasets have common properties with different data [6], [13]. The most apparent properties are the size of a dataset and the type of smells supported by the dataset. This paper aims to identify a standard scheme and architecture that code smells datasets can be fairly compared and ranked according to them. We investigate the current state of the available datasets for code smell detection using a systematic literature review (SLR). The anatomy of the code smell datasets is discussed from different viewpoints enabling us to identify the opportunities and pitfalls of the research in this area. Our systematic literature review is conducted to answer the following research questions:

---

[*] Corresponding author. Address: School of Computer Engineering, Iran University of Science and Technology, Hengam St., Resalat Sq., Tehran, Iran, Postal Code: 16846-13114.

- **RQ1**. *How many code smell datasets have been proposed by the software engineering community?*
- **RQ2**. *What are the common aspects of the code smells dataset anatomies?*
- **RQ3**. *What are the code smell dataset creation techniques and validation mechanisms?*
- **RQ4**. *Which software tools are mostly leveraged to automatically create code smells datasets?*
- **RQ5**. *Which programming languages, code smells, and code metrics are covered by existing datasets?*
- **RQ6**. *Which open-source or close-source software projects are widely used as data sources to create code smells datasets?*
- **RQ7**. *What are the publicly available code smell datasets?*
- **RQ8**. *How is the quality of the existing code smell datasets regarding different evaluation metrics?*
- **RQ9**. *What are the most comprehensive and adequate code smell datasets?*
- **RQ10**. *What are the limitations of the existing code smell datasets?*

A comprehensive search was performed about the code smell datasets on five digital libraries indexing the relevant publications of the field to find the answer to each research question. Our search has resulted in finding 2696 articles, of which 45 articles are of high quality and present new datasets. We compare the datasets according to different aspects, including size, supported smells, programming languages, and construction approaches. Our SLR indicates that while the field is growing and the changes are very dynamic, the existing code smell datasets suffer from many challenges. Most importantly, the small size, few types of code smell, high false-positive rate, and the lack of standard structure. We highlight the potential solutions to be considered in future research. To the best of our knowledge, this is the first systematic review of code smell datasets that helps researchers and practitioners to find the most appropriate datasets and improve them.

The remainder of the paper is organized as follows. Section 2 reviews the related SLRs on software smells and describes their difference from our proposed study. Section 3 outlines the research methodology of our SLR. The results of our findings on the primary studies are discussed in Section 4. Section 5 proposes a catalog with an in-depth review of the most notable code smell datasets. Section 6 discusses the challenge and opportunities in the area of code smell datasets. The threats to the validity of our SLR are described in Section 7. Finally, Section 8 concludes this paper and points out directions for future works.

## 2 RELATED RESEARCH

We found relatively few SLRs on code smells [2], [9]–[12], [14]. Most of them have been published in recent years, which denotes the importance and growing research in the field. However, none of them have studied the code datasets in detail. To the best of our knowledge, this paper is the first systematic literature review dedicated to code smell datasets in advance.

Caram et al. [11] provided a systematic mapping study to determine which methods, practices, techniques, and models are used when applying machine learning for code smell detection and prediction. The authors have identified 26 primary studies that used learning-based techniques for code smell identification. Bloaters [15] *i.e.*, long method, large class, primitive obsession, long parameter list, and data clumps have been studied in 35% of the papers. Genetic algorithms as the most commonly applied technique were used by 22.22% of the primary studies. A high level of redundancy has been reported regarding the smells addressed by each learning-based technique. In other words, most smells were detected by more than one algorithm. Feature envy was detected by 63% of the proposed techniques as the most common smell type detected automatically, while five out of the 25 analyzed smells have not been detected by any machine learning techniques. Regarding the F1 score the best average performance has been reported for the decision tree classifier in detecting middle-man and shotgun surgery smells, while random forest has an outstanding performance for the long parameter list. As a result, no machine learning technique is superior for all types of smells. The authors have found a lack of comparable results due to the heterogeneity of the data sources in the reported experiments.

Azeem et al. [2] proposed a systematic literature review on the use of learning-based techniques for code smell detection. The authors targeted four specific aspects related to how previous research conducted experimentations on code smell prediction models, including (i) which code smells have been considered, (ii) How the machine learning setups have been adopted, (iii) which types of evaluations strategies have been exploited, and (iv) what are performance claimed for the proposed machine learning models. Their analyses highlighted a number of limitations of existing studies as well as open issues that need to be addressed by future research. They also provided a list of actions that must be performed to make the research field more concrete, mainly prioritizing smells, configuring machine learning models, and preparing manually validated code smell datasets. In this paper, we list all available code smell datasets and describe their construction and validation mechanisms.

Trindade et al. [16] presented a systematic literature review to compile oracles for bad smells. The primary motivation of their study is the fact that bad smells have been widely studied, and many studies rely on bad smell oracles obtained by tools that have not yet been adequately proven to be precise. The oracles of bad smells available online have the following main characteristics. They involve a maximum of 29 software systems, varying from 86 to 17167 classes. Most of them rely on results provided by tools. They usually verify Large Class smell instances and provide their results in a spreadsheet. Their study indicates that researchers must address gaps related to code smell oracles. For example, just a few oracles accurately identify the methods where the bad smell is located. In addition, most oracles are indeed defined by employing tools, which is a serious threat to their validity. We expand the Trindade et al. [16] study to additional aspects of code smell datasets, mainly their construction and validation mechanisms.

Al-Shaaby et al. [17] systematically reviewed and analyzed the machine-learning approaches applied to code smell detection from different aspects, including smell types, learning algorithms, smell datasets, and software tools. They also reported a comparison between machine learning models used for code smell detection in terms of prediction accuracy and performance. Seventeen primary studies were selected and discussed in their SLR to answer the five research questions. The authors have concluded that the application of machine learning techniques to detect code smells is still a new area and needs further investigation. Therefore, more research efforts are required to facilitate the employment of machine learning techniques addressing the code smell prediction issues. We believe that many of these issues originate from the datasets, not the machine learning techniques used to predict.

Sobrinho et al. [14] conducted an extensive literature review on a huge body of knowledge from 1990 to 2017. They found that some smells are much more studied in the literature than others, and some of them are intrinsically interrelated (which). They give a perspective on how the research has been driven across time (when). They analyzed aims, findings, and respective experimental settings and observed that the variability of these elements might be responsible for some contradictory results on bad smells (what). Moreover, while bad smells of different types are generally studied together, only a tiny fraction of the studies have investigated the relationships between different smells (co-studies). The authors also mentioned that researchers have various interest levels in the subject, some of them publishing sporadically and others continuously (who). Their results show that the communities studying code clones or duplications and other types of bad smells are largely separated. The authors observed that some conferences and journals are more likely to disseminate knowledge on code clones while others follow a balanced distribution among all smells (where). We answer similar questions about the code smell datasets in this paper to shed light on the reasons behind the findings by Sobrinho et al. [14].

## 3   RESEARCH METHODOLOGY

We adapted the guidelines proposed for conducting SLR in software engineering [18]–[20] to identify, analyze, and assess the published literature about code smell datasets considering our research questions. Figure 1 shows the overall process of searching and selecting relevant publications. At first, we defined a search string containing relevant keywords to our SLR. Then, search the resultant query in the top five digital libraries indexing computer science literature massively, shown in Figure 1. Afterward, the article selection process is performed, and finally, a set of 45 articles are selected for a detailed review. We discuss each step in detail in the subsequent sections.

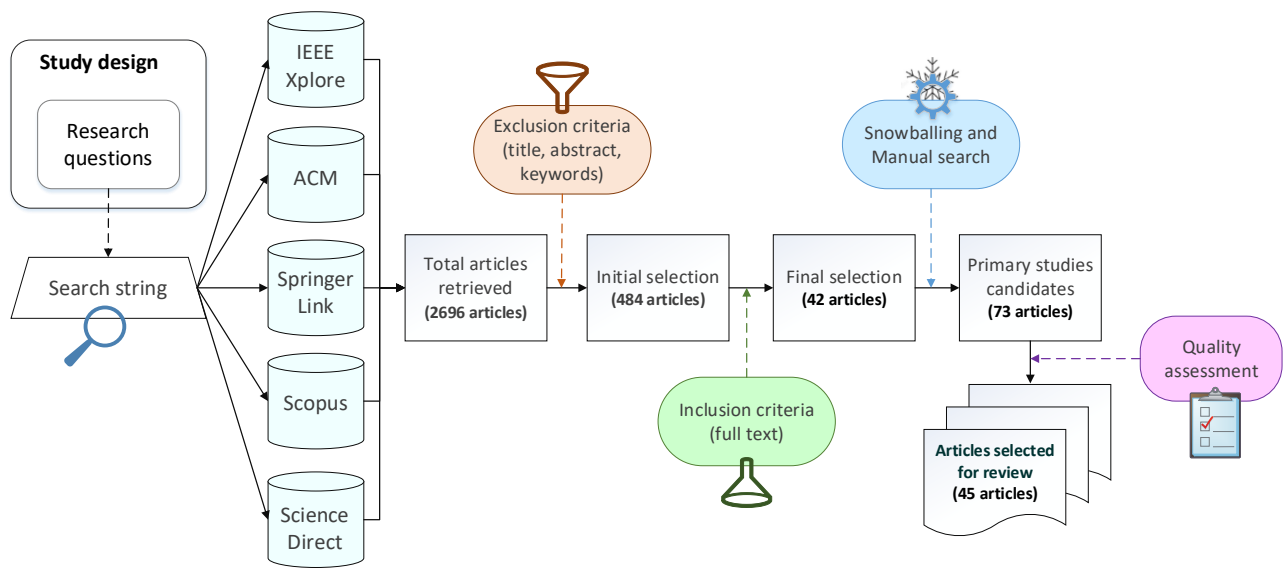

Figure 1. SLR process.

### 3.1   Constructing search string

Our search string is figured around three essential concepts, "source code," "smell," and "dataset," which appear in our research questions. The variation of these three keywords, combined with the Boolean operators, constitute our search string. We performed five steps recommended by Kitchenham and Charters [20] to find all relevant search terms and construct the required query string:

**Step 1**: We used the research questions in Section 1 to derive the main terms by identifying PICOC criteria [20], specifically the population, intervention, outcome, and context.

**Step 2**: We extracted and added related terms to the main terms as well as alternative spellings and synonyms of the main terms to our search string.

**Step 3:** We retrieved and verified the keywords in the related works incrementally and iteratively. Indeed, after performing an initial search, we checked the keywords in the most relevant articles to ensure adding any existing synonyms, spelling forms, and related words to our main terms used in the literature.

**Step 4:** We used the "OR" operator for concatenating the alternative spellings, synonyms, and related terms. Moreover, the "AND" operator was used for combining the main terms.

**Step 5:** We integrated the search string into a summarized form whenever it was required, according to the search engine's capabilities and limitations.

The results of each step are described as follows:

**Results for step 1:** As for the first step, the population, intervention, and outcome, were identified to find and organized the main terms. Our population is the code smells and anti-patterns appearing in the production code of software systems. The interventions are techniques, algorithms, tools, and datasets developed for code smells detection, identification, and prediction. The outcomes are different aspects of code smell datasets and databases introduced in either academic or industrial contexts, including the dataset labeling approach, structure, source of data, availability, and quality assessment. For instance, RQ6 contains the main terms related to our PICOC criteria when decomposed as follows: "*Which open-source or close-source* [*software projects*] **(outcomes)** *are widely used as data sources to create* [*code smells*] **(population)** [*datasets*] **(intervention)**? "

**Results for step 2:** The extracted synonyms, alternative spellings, and related terms of each main term are:

- Code smell: "bad smell" OR "smelly code" OR "anti-pattern" OR "anti pattern" OR "antipattern" OR "design smell"
- Detection: "detect" OR "identify" OR "identification" OR "predict" OR "prediction" OR "recognize"
- Software: "program" OR "metric"
- Dataset: "data set" OR "data-set" OR "benchmark" OR "oracle" OR "machine learning" OR "classification" OR "regression"

It is worth mentioning that anti-patterns and smells are near concepts such that software engineering researchers and practitioners often use them interchangeably [9]. We observed that some code smell datasets contain both the code smells and anti-patterns samples. Therefore, we decide to add the term "antipattern" to retrieve all related datasets in the field.

**Results for step 3:** The following keywords were extracted and added to our search terms after investigating the related research in step 3:

- Code smell: "design flaw"
- Detection: "refactor" OR "analysis" OR "empirical study"
- Software: "program" OR "metric" OR "maintenance"
- Dataset: "supervised learning" OR "unsupervised learning" OR "heuristic" OR "statistic"

**Results for step 4:** The combination of the extracted search terms with the help of the Boolean operators results in the final search string shown in the first row of Table 1.

**Results for step 5:** We optimized the query built upon our search string for each digital library due to specific formats and limitations, such as the query length imposed by their search engines. More precisely, we could not use the above search string with the ScienceDirect search engine because the engine has a limitation of accepting search strings including up to 8 logical operators. Therefore, we had to look for the most relevant keywords in our search string and remove the additional ones. Apparently, by restricting the logical operators the search results were applied to a broader number of articles and we had to look for the most relevant keywords in our search string and remove the additional ones. The search string used with the ScienceDirect library is shown in the second row of Table 1.

Table 1. Search string used to query each digital library.

| Repository | Search string |
| --- | --- |
| IEEE Xplore, ACM, SpringerLink, and Scopus | ((("code smell" OR "bad smell" OR "smelly code" OR "anti-pattern" OR "anti pattern" OR "antipattern" OR "design smell" OR "design flaw") **AND** ("detection" OR "detect" OR "identify" OR "identification" OR "predict" OR "prediction" OR "recognize" OR "empirical study" OR "analysis" OR "refactor") **AND** ("program" OR "software" OR "metric" OR "maintenance") **AND** ("dataset" OR "data set" OR "data-set" OR "benchmark" OR "oracle" OR "machine learning" OR "supervised learning" OR "unsupervised learning" OR "classification" OR "regression" OR "heuristic" OR "statistic")) |
| ScienceDirect | ("code smell" OR "antipattern" OR "design smell") **AND** ("detect") **AND** ("code" OR "design") **AND** ("dataset" OR "data set" OR "machine learning") |

## 3.2 Resources to be searched

Choosing the proper resources to search for relevant literature plays a significant role in an SLR. We selected the five well-known digital libraries which mainly index computer science publications as resources to search for all the available literature related to our research questions: (1) IEEE Xplore digital library (*http://ieeexplore.ieee.org*), (2) ACM digital library (*https://dl.acm.org*), (3) SpringerLink (*https://link.springer.com*), (4) Scopus (*https://www.scopus.com*), (5) ScienceDirect (*http://www.sciencedirect.com*).

The five digital libraries contain a large portion of publications in software engineering and machine learning, including journal articles, conference proceedings, book chapters, and books. Therefore, we ensure that all related works published in code smells and the dataset area are found and retrieved.

### 3.3 Article selection process

The article selection consists of five steps, shown in Figure 1. In the initial search of the repositories for our search string, no restrictions were placed on the publisher, publication type, year of publication, or other items. As a result, a set of 2696 resources were found and retrieved in our initial search, indicating a pretty large number of publications in this area.

We used five exclusion criteria (ECs) in the second step to filter out the initial set of articles and remove irrelevant studies. A Python script [21] was developed to automatically apply exclusion criteria where possible. Thereafter, the third author (M.Sc. in software engineering) checked the exclusion criteria on the remained articles considering each resource's title, abstract, keywords, and the manuscript text, where it was required. The results were double-checked by the first author (Ph.D. candidate in software engineering). In case of disagreement in selecting or excluding an article, the second and fourth authors were asked to make the final decision based on the response provided by the first and three authors. It led to the discarding of 2212 resources, resulting in 484 papers to proceed in the next steps. The applied exclusion criteria are as follows:

- **EC1**: *Duplicated resources*: Some articles were indexed by more than one digital library. We kept one copy of each duplicated paper.
- **EC2**: *Non-English resources*: We removed manuscripts written in languages other than English. We observed that none of them provided a new public dataset, including code smell.
- **EC3**: *Non-primary studies*: All secondary and tertiary studies were removed. We discussed them as related work to our SLR in Section 2.
- **EC4**: *Full-text availability*: Articles whose full-text was not available or less than two pages were removed.
- **EC5**: *Irrelevant papers*: Articles focused on other software engineering topics rather than code smells. Our search string terms such as "code smell" and "dataset" only appeared in the related works section of these articles, or each term appeared in a different article section. For instance, the word "dataset" did not point to a code smells dataset in the paper.

In the third step, two inclusion criteria were applied to find the relevant papers appropriate for a detailed study. We selected the articles that dedicated a section to describing their "code smell dataset." In addition, we chose those articles that implicitly explain their "code smell dataset" in the article's main text. To this aim, the contents of the papers obtained in the second step were jointly reviewed by the first and third authors, respectively, a Ph.D. candidate and an M.Sc. graduate in software engineering. We extracted a set of information, including dataset creation and validation process, studied projects, code smells, number of smelly and non-smelly samples from each paper during the review, and saved them in a Microsoft Excel file. The completed Excel file was further used in our manual analysis to answer the research questions described in Section 1. Each selected article was also investigated and verified by the second and fourth authors of the paper, who have a Ph.D. in computer science and are experts in software engineering, to ensure the relevancy and usefulness of the selected papers and obtained information. As shown in Figure 1, 42 articles were picked systematically at the end of this step.

In the fourth step, backward and forward snowballing [22] was carefully performed on the candidate papers by the first and third authors. The snowballing process resulted in 31 new articles. It is worth noting that we included all types of resources, *i.e.*, journals, conferences, workshops, and technical reports to achieve a collection of relevant papers that are as comprehensive as possible.

Finally, in the last step, the quality assessment is carefully performed on 73 papers by answering the questions about the quality of the paper and assessing the score computed for each article regarding these questions. The details of the study quality assessment are described in Section 3.4.

### 3.4 Quality assessment

We used the following checklist to assess the credibility and thoroughness of the selected publications in the study quality assessment step.

- **Q1**: *Is the dataset related to code smell?*
- **Q2**: *Is the dataset new?*
    - Q2.1: *If not, do the authors add any new features to the dataset?*
- **Q3**: *Do the authors provide helpful information for their datasets?*
    - Q3.1: *Do the authors mention the dataset construction and validation mechanisms?*
    - Q3.2: *Do the authors mention the tool(s) used to create or validate the dataset?*
    - Q3.3: *Do the authors mention the source projects of their dataset?*
    - Q3.4: *Do the authors mention the number of each code smell in the dataset?*
    - Q3.5: *Do the authors mention the amount of effort used to prepare the dataset?*
- **Q4**: *Is the dataset created by a novel approach, or does it have a considerable advantage?*

The answer to each question in our checklist was marked with "Yes," "No," or "Partially" for each primary study in the candidate list of studies. The subquestions were evaluated to determine the partial answers. The term "Partially" was used when some of the subquestions in the quality assessment checklist are answered with "Yes" and the others are answered with No. In such cases, the main question was marked as "Partially". We then scored the answers based on the following rules: "Yes" = 1, "No" = 0, ad "Partially" = 0.5. For each candidate's primary study, its quality score was computed by summing up the scores of the answers to all four questions.

The scoring process was performed by the first and third authors, who jointly evaluated each article in the set of 73 candidate studies and used consensus to determine the final score. We categorized the quality level into High (score = 4), Medium ($2 \leq$ score $< 4$), and Low (score $< 2$). The articles whose scores belonged to the high and medium levels were selected for in-depth analysis as our final primary studies. Articles that used other researchers' or their previous datasets or utilized datasets unrelated to code smell, or did not provide helpful information for their datasets received low scores and were removed from our repository. In the same way, articles that were created with a similar approach to those in our collection received a low score and were eliminated. For instance, two articles [23], [24] are similar to S22, S4, and S5.

Our search ended up with a total of 45 papers that highly contributed to the area of code smells datasets and validation mechanisms. We could retrieve only 25 datasets from the internet, indicating a relatively low number of publicly available code smells datasets. We analyzed all datasets manually or by simple Python scripts to extract the required information. Objective information extracted from the primary studies was also checked by three independent M.Sc. students in software engineering in addition to the authors to ensure the correctness of the results reported in Section 4. The Microsoft Excel file containing the data extracted during the article selection process is publicly available on Zenodo in reference [25].

## 4   FINDINGS AND RESULTS

We describe our findings from the objective investigation of the primary studies found by the research methodology described in Section 3. The section begins with an overview of the reviewed articles to answer RQ1.

### 4.1   Overview

This section aims to answer our first research question, *how many code smell datasets have been proposed by the software engineering community?* To answer RQ1, we investigate the frequency and diversity of the proposed code smell datasets in existing primary studies. Table 2 shows the list of publications contributing to the code smells datasets obtained by our proposed resource selection process. We observe that 26 of 45 selected resources belong to conference articles, 18 papers belong to journal articles, and one resource is a book chapter. It concludes that the topic of the code smell dataset is covered by various publication types. Regarding the title of primary studies, only four papers, S3, S8, S10, and S11, directly point out the dataset term in their title. In other words, a few articles are dedicatedly studied code smell datasets. Table 3 shows the initial and final number of retrieved articles from each digital library. Google Scholar is not used as a primary library for searching and extracting resources. However, we found two articles that previous libraries have not indexed in the manual search and snowballing step. It is observed that the IEEE Xplore digital library hosts most (23 out of 45) of the primary studies about code smell datasets.

Table 2. Articles investigated and reviewed in our SLR.

| Study | Title | First author | Type | Ref. |
|---|---|---|---|---|
| S1 | On the diffuseness and the impact on maintainability of code smells: a large-scale empirical investigation | F. Palomba | Journal | [26] |
| S2 | A large-scale empirical study on the lifecycle of code smell co-occurrences | F. Palomba | Journal | [27] |
| S3 | Landfill: an open dataset of code smells with public evaluation | F. Palomba | Conference | [28] |
| S4 | Comparing and experimenting machine learning techniques for code smell detection | F. Fontana | Journal | [29] |
| S5 | Code smell severity classification using machine learning techniques | F. Fontana | Journal | [30] |
| S6 | Code smell prediction employing machine learning meets emerging Java language constructs | H. Grodzicka | Book chapter | [31] |
| S7 | Detecting code smells using machine learning techniques: are we there yet? | D. Di Nucci | Conference | [3] |
| S8 | The technical debt dataset | V. Lenarduzzi | Conference | [32] |
| S9 | Code smell detection using multi-label classification approach | T. Guggulothu | Journal | [33] |
| S10 | MLCQ: industry-relevant code smell data set | L. Madeyski | Conference | [34] |
| S11 | Using code evolution information to improve the quality of labels in code smell datasets | Y. Wang | Conference | [35] |
| S12 | Comparing heuristic and machine learning approaches for metric-based code smell detection | F. Pecorelli | Conference | [36] |
| S13 | A support vector machine based approach for code smell detection | A. Kaur | Conference | [37] |
| S14 | Experience report: evaluating the effectiveness of decision trees for detecting code smells | L. Amorim | Conference | [38] |
| S15 | Context-based code smells prioritization for prefactoring | N. Sae-Lim | Conference | [39] |
| S16 | Bad-smell prediction from software design model using machine learning techniques | N. Maneerat | Conference | [40] |
| S17 | Evaluating the accuracy of machine learning algorithms on detecting code smells for different developers | M. Hozano | Conference | [41] |
| S18 | Competitive coevolutionary code-smells detection | M. Boussaa | Conference | [42] |
| S19 | A machine learning based ensemble method for anti-patterns detection | A. Barbez | Journal | [43] |
| S20 | Finding bad code smells with neural network models | D. Kim | Journal | [44] |
| S21 | Evaluation of machine learning approaches for change-proneness prediction using code smells | K. Kaur | Conference | [45] |
| S22 | SMURF: a SVM-based incremental anti-pattern detection approach | A. Maiga | Conference | [46] |
| S23 | BDTEX: a GQM-based Bayesian approach for the detection of anti-patterns | F. Khomh | Journal | [47] |
| S24 | Reducing subjectivity in code smells detection: experimenting with the long method | S. Bryton | Conference | [48] |
| S25 | Adaptive detection of design flaws | J. Kreimer | Journal | [49] |
| S26 | Classification model for code clones based on machine learning | J. Yang | Journal | [50] |
| S27 | Can I clone this piece of code here? | X. Wang | Conference | [51] |

**A systematic literature review on the code smells datasets and evaluation techniques** *Zakeri-Nasrabadi et al.*

| Study | Title | First author | Type | Ref. |
|---|---|---|---|---|
| S28 | An immune-inspired approach for the detection of software design smells | S. Hassaine | Conference | [52] |
| S29 | Tracking design smells lessons from a study of God classes | S. Vaucher | Conference | [53] |
| S30 | A Bayesian approach for the detection of code and design smells | F. Khomh | Conference | [54] |
| S31 | Predicting maintainability of open-source software using gene expression programming and bad smells | S. Tarwani | Conference | [55] |
| S32 | An exploratory study of the impact of anti-patterns on class change- and fault-proneness | F.Khomh | Journal | [56] |
| S33 | DÉCOR: A method for the specification and detection of code and design smells | N. Moha | Journal | [57] |
| S34 | Developer-driven code smell prioritization | F. Pecorelli | Conference | [58] |
| S35 | Software code smell prediction model using Shannon, Rényi, and Tsallis entropies | A. Gupta | Journal | [59] |
| S36 | Application of machine learning algorithms for code smell prediction using object-oriented software metrics | M. Agnihotri | Journal | [60] |
| S37 | Toward a smell-aware bug prediction model | F. Palomba | Journal | [61] |
| S38 | Detecting bad smells with machine learning algorithms: an empirical study | D. Cruz | Conference | [62] |
| S39 | Beyond technical aspects: how do community smells influence the intensity of code smells? | F. Palomba | Journal | [63] |
| S40 | An empirical study of the performance impacts of android code smells | G. Hecht | Conference | [64] |
| S41 | Detecting bad smells in source code using change history information | F. Palomba | Conference | [65] |
| S42 | An exploratory study of the impact of code smells on software change-proneness | F. Khomh | Conference | [56] |
| S43 | Detection of shotgun surgery and message chain code smells using machine learning techniques | T. Guggulothu | Journal | [66] |
| S44 | Deep learning based code smell detection | H. Liu | Journal | [67] |
| S45 | Detecting code smells using deep learning | A. Das | Conference | [68] |

Table 3. Number of retrieved resources by each digital library search engine

| Library | Initial number | Final number | Journal/ book chapter | Conference |
|---|---|---|---|---|
| IEEE Xplore | 65 | 23 | 4 | 19 |
| ACM | 821 | 4 | 0 | 4 |
| SpringerLink | 1329 | 8 | 6 | 2 |
| Scopus | 301 | 1 | 1 | 0 |
| ScienceDirect | 180 | 5 | 5 | 0 |
| Google Scholar | 0 | 4 | 3 | 1 |
| Total | 2696 | 45 | 19 (42%) | 26 (58%) |

**RQ1**: *How many code smell datasets have been proposed by the software engineering community?*

**Summary for RQ1**: *A total of 45 primary studies exist that contain a dataset, related to code smells and anti-patterns. The title of most published papers does not directly point to the dataset. Only four out of 45 papers have dedicatedly discussed code smell datasets. Nearly 58% of primary studies have been published as conference papers, and the reaming 42% are journal articles, indicating the importance of the topic. IEEE Xplore has indexed and published 51% of all studies related to code smell datasets.*

### 4.2 Code smell datasets classification

It is essential to crafting an abstract model to categorize and compare code smell datasets in a standard and fair scheme. This section answers our second research question, "*what are the common aspects of the code smell dataset anatomies?*" To this aim, the keywords of our search string that appeared in RQs 3 to 8 were extracted to find the features of code smell datasets for which we investigated the primary studies. The investigation of these features in the primary studies indicated that existing code smells datasets could be categorized and compared from five orthogonal aspects, including labeling, structure, data source, availability, and quality. After that, the first three authors manually analyzed the data in the Microsoft Excel file [25] described in Section 3 to specify the categories in each aspect and find the number of studies in each category.

Figure 2 shows the classification of the code smell datasets regarding the extracted aspects and categories. Each category includes several subcategories organized in a hierarchical form. The numbers inside the parenthesis in the leaves of the classification tree denote the number of existing datasets in that subcategory. We organized the technical review of code smell datasets introduced by the primary studies according to the proposed classification. The subsequent sections discuss each aspect in detail to provide answers to other research questions mentioned in Section 1.

The most important aspect of code smells datasets in our classification is the 'labeling' aspect which describes the oracle creation and validation process used for each dataset. Three oracle creation techniques and three oracle validation mechanisms are observed regarding the dataset labeling process. Figure 2 shows that the oracles of most code smell datasets are created automatically and then validated manually by experts or no validation has been performed. Section 4.3 discusses the labeling process of existing datasets in detail.

The second aspect that distinguishes code smell datasets is 'structure'. We identified five subcategories related to this aspect including supported languages, smell types, severity levels, features, and instance ratio. Regarding programming languages, 43 out of 45 primary studies have proposed a code smell dataset for Java programs. Therefore, the analytical results in our study are limited to the concept of code smells and metrics in object-oriented programming languages. Section 4.4 investigates the structure of the code smell dataset in detail.

The third aspect that differentiates code smell datasets is the 'source of data' describing the software systems used as benchmark projects to create dataset samples. We distinguish between open-source, industrial, and academic projects as well as their combination when discussing the data source of code smell datasets. As shown in Figure 2, the majority of datasets have been built based on open-source software projects. The details about software systems used in the code smell dataset are discussed in Section 4.5.

The fourth aspect that distinguishes code smell datasets is the 'availability' of a proposed dataset. The availability is an essential factor for researchers and practitioners who want to use or contribute to them. It is observed that more than half of code smell datasets are outdated or not available publicly highlighting the need for public datasets in the field. The publicly available code smell datasets are discussed in Section 4.6. Finally, the fifth aspect in our classification is 'quality' describing which evaluation metrics have been used to assess the correctness of the proposed dataset. Section 4.7 compare the quality of existing code smell datasets regarding the reported evaluation metrics and review the advantages and disadvantage of each dataset.

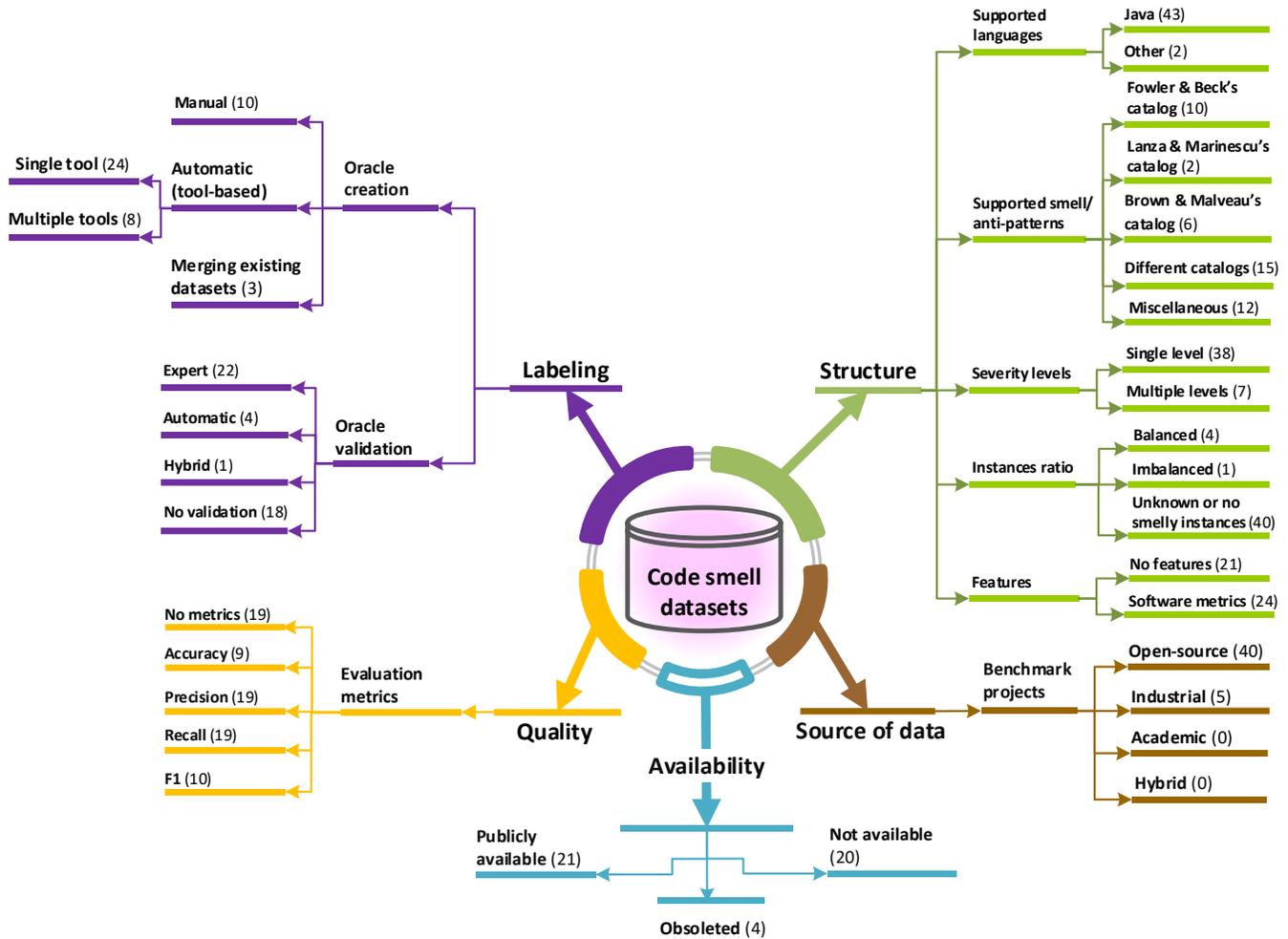

Figure 2. Classification of code smell datasets.

> **RQ2**: *What are the common aspects of the code smells dataset anatomies?*
>
> **Summary for RQ2**: *The deep investigation of the proposed code smell datasets reveals the existence of five orthogonal aspects as a basis for classifying datasets. Existing code smell datasets are classifiable according to their labeling process, structure, source of data, availability, and quality. Regarding the dataset labeling process, three oracle creation techniques and three oracle validation mechanisms are observed. Machines (automatic approaches) are competing with humans with a ratio of 35 to 30 in code smell oracle creation and validation processes. Regarding the structure, code smell datasets are distinguishable by supported languages, smell types, severity levels, instances ratio, and additional features. The source of data for creating code smell datasets come from open-source, industrial, or academic software projects. Regarding availability, there are public, non-public, and obsoleted code smells datasets. Finally, the quality of the code smell datasets is evaluated based on four metrics.*

### 4.3    Code smell datasets' labeling

The third research question, *what are the code smell dataset creation techniques and validation mechanisms*, is answered by analyzing the labeling process of code smell datasets in our primary studies. According to Figure 2, the labeling process in code smell datasets includes two phases of oracle creation and oracle validation. During the oracle creation, a label is assigned to each program entity (*e.g.*, method or class)

A systematic literature review on the code smells datasets and evaluation techniques                     *Zakeri-Nasrabadi et al.*

specifying its smell type. Entities with no labels are considered non-smelly or smell-free instances. In the oracle validation phase, assigned labels are checked to ensure their correctness and fix the false ones. The labeling process highly affects the reliability and correctness of a code smell dataset regardless of its structural properties, which mainly affects the development of code smell detection tools. Due to numerous entities in the source code of real-world programs, various oracle creation and validation methods have been presented.

### 4.3.1 Oracle creation approaches

Table 4 summarizes the contributions proposed in the primary studies regarding the construction and validations of their code smell datasets. It is observed that researchers have used three approaches to create oracles: manual, automatic (tool-based), and merging. Following are descriptions of each oracle creation approach:

**Manual**. The manual approaches use experienced software developers and practitioners (called experts) to recognize code smells, such as those proposed in S10 and S17. Creating a large and quality dataset can be a time-consuming and expensive process, threatened by the experts' opinions and knowledge [8]. The results of manual labeling are typically validated during the oracle validation process by another group of experts to reduce the bias caused by the first group.

**Automatic (tool-based)**: The automatic approaches use existing code smell detection and refactoring tools and do not rely on human experts. Automatic creation of a large code smell dataset, *e.g.*, S8, S11, required less effort than the manual approach. However, a fully automated labeling approach invariably leads to many false-positives samples (refer to Section 4.3.2) [69]. Moreover, the type of smell is limited by existing tools. In other words, automatic approaches cannot be used to identify new types of code smells. Human experts often involve in the oracle validation phase to identify and remove false-positive samples as much as possible. S1 and S4 have employed human experts to filter out false-positive samples. The tool-based creation approach used by Boussaa et al. in S18 differs from the other automatically created datasets. Instead of detecting code smells using tools, they created prototypes of artificial code smells. Since the artificially generated smells are different from the actual ones, they have added some real smelly instances to their dataset to improve the naturalness of the data and achieve better results. The construction approach of some datasets, such as S13, has not been mentioned clearly.

**Merging**: In this technique, existing code smell datasets are combined to form a new dataset with more samples and smell types compared to the existing ones. During the merge process, some properties such as the order, place, and structure of samples may be changed to enhance representation. Moreover, a new round of validation process may be applied to entire samples to improve dataset quality. Authors in S7, S9, and S16 have used merging to create new code smell datasets. In S7 and S9, Fontana's dataset (S4) files have been merged to address some problems described in Section 5. In S16, the authors have collected seven datasets from the previous works, which offer 27 design model metrics and seven code smells. We believe that merging is a promising approach to achieving a larger dataset while it requires proper validation to access a high-quality dataset.

### 4.3.2 Oracle validation mechanisms

After constructing the code smell dataset, an important step is to validate the dataset. The validation process mainly affects the results obtained by computing evaluation metrics on every code smell detection or prediction approach. The third column of Table 4 indicates the validation mechanism associated with each primary study. We found that the code smell datasets are primarily validated in three methods: Manual validation by one or several experts, automatic validation, and a combination of them. Moreover, some authors have not validated their datasets. Studies without a clear validation mechanism are denoted with a "No validation" term in the third column of Table 4. It is observed that 18 out of 45 datasets provide no validations.

Experts are people with significant hands-on experience in detecting, classifying, and fixing code smells in software systems. Most manually validated code smell datasets have been validated by only one expert, threatening the reliability of available datasets and increasing the bias toward one person's opinions. Our findings show that experts who evaluated the code smell datasets are the authors of the datasets themselves, trained students, and professional developers. Mello et al. [70] have shown that reviewers' collaboration significantly increases the precision of smell identification. They also have shown that having previous knowledge of the reviewed module does not affect the precision of reviewers with higher professional backgrounds [70]. Therefore, it is expected to involve more human evaluators even if they are not highly expert developers when creating new code smell datasets.

Automatic validation is used when dataset creators apply an automated mechanism to evaluate a dataset, *e.g.*, using unsupervised learning [66] or the history of changes [35] in the refactoring process and different versions of software. Both types of manual and automatic validations are accompanied by a voting mechanism when three or more validators (experts or algorithms) are involved in making the final decision about smelly and non-smelly samples and the type of smell. The "voting" term has been added to the validation mechanisms in Table 4 for datasets validated by three or more validators. Hybrid validation had been only used by the authors of S11 [35]. Wang et al. in S11 have invented a hybrid technique based on code evolution information. They have automatically evaluated samples by analyzing two versions of a refactored software. Only smelly samples in the first version that are not smelly in refactored version are considered as the true sample of a code smell. Wang et al. have also performed manual evaluations to ensure the reliability of the results. As a result, regarding the quality of the validation process, S11 is one of the prominent studies. In contrast, many studies, such as S7-S9, S13-S16, and S20-S22, have not performed any validation activity. The validation process should be taken into account even for datasets that are created by merging existing ones since the base datasets are often validated under different conditions and are associated with different qualities.

One may argue that the dataset validation mechanisms are primarily related to the dataset construction approaches. However, it is not true in general. For example, tool-based created datasets are often evaluated manually. Code smell dataset contributors need to distinguish between oracle creation and validation processes when introducing a new dataset in this field. The clarification of these concepts and serious attention to the validation process in the future make the empirical results more realistic than the current ones.

Table 4. Code smell datasets creation approaches and validation mechanisms.

| Study | Labeling approach | Validation mechanism |
|---|---|---|
| S1 | **Tool-based**: Using a simple detection tool to extract code smell candidates and then two authors validate them manually. Finally, they performed an open discussion to resolve possible conflicts and consensus on the detected code smells to ensure high recall. | Expert |
| S2 | **Tool-based**: A simple tool that discarded the classes/methods which surely do not contain code smells has been used. Then, two master students (*i.e.*, the inspectors) individually analyzed and classified code elements of each system as true positive or false positive for a given smell. All the instances positively classified by both inspectors have been considered as real smells. The inspectors opened a discussion to resolve the disagreement and make a shared decision for the other instance. | Expert |
| S3 | **Manual**: The first author manually detected the smelly instance, and then another author validated the produced oracle to verify the results. | Expert |
| S4 | **Tool-based**: Using several smell detection tools as advisors. The oracle evaluation was performed by three M.Sc. students trained explicitly for the task. The students independently studied the code smell definitions and held a two-hour discussion about their opinions. | Expert |
| S5 | **Tool-based**: Using several smell detection tools as advisors. The labeling process is also supported by graphical code representations, like dependencies, calls, and hierarchy graphs. Neither the values of software metrics nor the number of advisors suggesting an instance were available during the evaluation to avoid biases. Three M.Sc. students performed the labeling process after being trained both theoretically and practically for the task. | Expert |
| S6 | **Tool-based**: The author's tool, JavaMetrics, was used for the initial filtering of samples from the dataset, helping authors to conduct an in-depth analysis of the selected samples based on source code metrics. Code smell labeling was performed by the authors, 4th-year software engineering students, and developers with approximately one year of professional experience. | Expert |
| S7 | **Merging**: The authors have merged Fontana's dataset files (S4) to reduce the balancing rate and make a dataset with more than one type of smell. | No validation |
| S8 | **Tool-based**: The authors have cloned the projects' repositories and iterated on each commit using PyDriller [71]. For each commit, the following actions are performed:<br>• The commit information in the GitLog is retrieved using PyDriller [71].<br>• The refactorings are classified using RefactoringMiner [72].<br>• The code is analyzed with SonarQube [73] using the default quality model (Sonar way) to collect technical debt information<br>• Code smells and anti-patterns are detected with Ptidej [74]. | No validation |
| S9 | **Merging**: The authors have merged Fontana's dataset files (S4) to make a multilabel dataset. | No validation |
| S10 | **Manual**: The dataset provides unique and detailed insights related to the professional and academic backgrounds of the reviewers. All of the reviewers involved in the code smell assessment are actively employed in the software development industry. The majority of samples are gathered by developers that are neither students nor researchers. | Expert |
| S11 | **Tool-based**: The authors have used code evolution information to detect false positive instances and improve the quality of labels. | Expert and automatically |
| S12 | **Tool-based**: The authors have used DÉCOR [57] to detect code smells and then manually evaluate the samples. | Expert |
| S13 | **Manual**. Training dataset (TDS) = {$C_j$, where j=1, 2, …, n} and $C_j$ is a set of classes obtained from object-oriented systems. ∀j, $C_j$ is marked as smelly or not. Object-orient metrics have been computed for each class of TDS which are used as an attribute $A_j$ for each class of training dataset. An SVM classifier is used to detect the new existence of code smells in the dataset. | Expert |
| S14 | **Tool-based**: The information about smells in a class has been derived from the S32 dataset. The CKJM [75] and POM [76] tools were used to calculate 18 and 44 metrics, respectively, for each class. Both tools calculate some metrics, but they decided to keep both versions since each tool calculates metrics differently and thus, produces different values. It is essential to observe that some code smells are more related to some metrics than others. Therefore, the models used for code smell detection give more importance to some metrics than others. | No validation |
| S15 | **Tool-based**: The authors have used the Infusion tool [77] to detect code smells. | No validation |
| S16 | **Merging**: The authors have collected seven datasets from the previous works, which offer 27 design model metrics and seven code smells. | No validation |
| S17 | **Manual**: The authors have built a dataset based on the input of 40 experienced developers, which classified 15 code snippets according to their perspectives. They have used detailed data containing instances of 4 code smell types manually validated by 40 developers. The authors have also built a dataset containing examples of non-smelly code snippets according to the developers' input. | Expert |
| S18 | **Tool-based**: The artificial code-smell examples are generated based on the notion of deviance from well-designed code fragments. Indeed, since CCEA [78] generates "artificial" code smell examples thus, only a few manually collected code smells are required to achieve good detection results. This approach reduces the effort required by developers to inspect systems to produce code smell examples. | Automatically |
| S19 | **Tool-based**: First, the output of three detection tools (HIST [65], [79], Incode [80], and JDeodorant [81]) is merged while adjusting their detection thresholds to produce the number of candidates per system proportional to the systems sizes. Second, three different groups of people manually checked each candidate of this set: (1) the authors of the paper, (2) nine M.Sc. and Ph.D. students, and (3) two software engineers. They asked respondents to report their confidence in the qualitative range of strongly approve, weakly approve, and strongly disapprove. To avoid any bias, none of the respondents was aware of the origin of each candidate. The candidate is considered a smell if the mean weight of the three answers reported for it is greater than 0.5. | Expert voting (weighted) |
| S20 | **Tool-based**: The authors have used their own tool, Code smell detector, to detect whether or not the classes in the Java projects are smelly based on source code metrics. | No validation |
| S21 | **Tool-based**: The authors have used JDeodorant [81] to extract non-exception handling smells and Robusta [82] to collect exception handling smells. | No validation |

**A systematic literature review on the code smells datasets and evaluation techniques** — *Zakeri-Nasrabadi et al.*

| Study | Labeling approach | Validation mechanism |
|---|---|---|
| S22 | **Manual**: A set of classes $C_i$ has been derived from an object-oriented system that constitutes the training dataset. $\forall i: C_i$ is labeled as Smelly class (*e.g.*, blob) or not. A classifier has been trained on the dataset and used to identify code smells in new samples. After manually validating newly detected smelly samples, the correct ones were added to the training set. | Expert |
| S23 | **Manual**: The authors have asked four undergraduate students and three graduate students to identify occurrences of the three anti-patterns in the two programs. They independently combined the students' votes using majority voting. If at least three of the five students/pair considered a class an anti-pattern, they tagged it as a true occurrence, *i.e.*, an instance of the anti-pattern. | Expert voting |
| S24 | **Manual**: In this study, all three authors separately played the role of experts. Each method was independently inspected by each expert. They have only considered a method to be a long method in cases where they had a full match. | Expert |
| S25 | **Tool-based**: The authors' own tool, IYC (it is your code), suggests potential design flaws then the user decides whether a flaw exists. Then, manually validated samples prepare an initial training set to train a decision tree. The learned model is then used to detect new instances, validate them manually, and add them to the training set. | Expert |
| S26 | **Tool-based**: A clone detection tool detects a set of clones in the source code. The user marks some of these clones as true or false clones according to her/his judgment and then submits these marked clones to FICA (filter for the individual user on code clone analysis) as a profile. FICA records the marked clones in its own database. FICA ranks other unmarked clones based on the result of machine learning, which predicts the probability that each clone is relevant to the user. The user can adjust the marks on code clones and resubmit them to FICA to obtain a better prediction. | Expert |
| S27 | **Tool-based**: First, the authors have used a clone detector to identify several cloning operations performed in the version histories of existing software projects. Second, for each cloning operation acquired in the first step, they have determined the values of the 21 features of the cloning operation and whether it is harmful or harmless to form a training instance. Third, they have constructed the Bayesian network based on the training instances. | No validation |
| S28 | **Tool-based**: The oracles were manually created by analyzing the two systems used in the experiments. Three of the authors independently re-validated the publicly available data in S30 and S33 to reduce the risk of classification errors. A candidate smell was classified as an actual smell when two out of three authors classified it as a smelly instance. | Expert voting |
| S29 | **Manual**: The authors have asked two undergraduate students and two graduate students to detect occurrences of God class in the two systems. A pair of undergraduate students performed the task together. | Expert voting |
| S30 | **Manual**: The authors have asked two undergraduate students and two graduate students to detect occurrences of the Blob in the two programs. The student opinions have been independently combined such that if at least two of the three students/pair considered a class smelly, tagged as a smelly occurrence. | Expert voting |
| S31 | **Tool-based**: Eleven code smells have been identified in every class with the help of two tools, JDeodorant [81] and Robusta. | No validation |
| S32 | **Tool-based**: The authors have used their previous approach, DÉCOR (defect detection for correction) [57], to specify and detect anti-patterns. | No validation |
| S33 | **Tool-based**: The authors have automatically applied the detection algorithms (DÉCOR [57]) on models of systems to detect suspicious classes. Detection algorithms may be applied in isolation or batch. | No validation |
| S34 | **Tool-based**: The authors have built an automated mechanism that fetches daily commits from the repositories to a local copy. This allowed them to generate the list of classes modified during the workday. At this point, they performed the actual smell detection. The authors used DÉCOR [57] to identify instances of the blob, complex class, and spaghetti code and used HIST [65], [79] to detect shotgun surgery. Afterward, they manually double-checked the smelly classes given by the automated tools to discard possible false positives. Finally, they sent emails to the original developers to ask (i) whether s/he actually recognized the presence of a code smell and (ii) if so, rate its criticality using a Likert scale from 1 (very low) to 5 (very high). | Expert |
| S35 | **Tool-based**: The data consisting of six bad smells are extracted for seven official releases of the Apache Abdera project using the Robusta [82] smell detection tool. | No validation |
| S36 | **Tool-based**: Four code smells, feature envy, dispersed coupling, refused parent bequest, and God class, were identified using the Eclipse plugin JSpIRIT [83]. | No validation |
| S37 | **Tool-based**: The authors have relied on the smells detected by JCodeOdor [84] because it has been empirically validated, demonstrating good performances in detecting code smells and detecting all the code smells considered in the empirical studies. In addition, JCodeOdor [84] computes the value of the intensity index on the detected code smells. | No validation |
| S38 | **Tool-based**: The authors have combined the results of five automatic detection tools to create bad smells oracles. They applied three detection tools for each bad smell and computed an agreement voting between their results. An entity (class or method) is considered smelly if two or more tools detect it. | Automatically voting |
| S39 | **Tool-based**: To collect smell instances, the authors have selected DÉCOR [57] because it has been employed in previous investigations on code smells, demonstrating exemplary performance in terms of precision, recall, and scalability. | No validation |
| S40 | **Tool-based**: The authors have detected the three smells in the projects by performing a static analysis using the Paprika tool [85]. They have obtained a list of methods and classes concerned with the three code smells. Then, they manually corrected each smell. | Expert |
| S41 | **Manual**: A M.Sc. student manually identified instances of the five considered smells in each system's snapshots. Starting from the definition of the five smells reported in the literature, the student manually analyzed each snapshot's source code looking for instances of those smells. Clearly, for smells having an intrinsic historical nature, he analyzed the changes performed by developers on different code components. A second M.Sc. student validated the produced oracle to verify that all affected code components identified by the first student were correct. | Expert |
| S42 | **Tool-based**: The authors have used their previously proposed tool, DÉCOR [57], to specify and detect code smells. | No validation |
| S43 | **Tool-based**: The researchers have assigned class instances labels with the help of detection rules proposed in the literature (JCodeOdor [84]). The method instances affected (positive) by the above rules have been compared with the formed clusters to validate the instances. If an instance produces the same cluster as its label, it is considered smelly. | Automatically |
| S44 | **Tool-based**: Applying refactoring, $ar$, to well-designed applications would change their well-designed internal structures. As a result, the refactoring leads to bad or suboptimal design, *i.e.*, code smells. The resultant smells should be resolved by applying another software refactoring, $ur$. Indeed, refactoring, $ur$, does nothing except undo the smell-introducing refactoring, $ar$. An example of smell introducing refactoring is to move a method from a class, $sc$ (where the method should be placed) to another class, $tc$. The move method operation, in this case, results in a feature envy smell. The smell is resolved by another move method operation that moves the method from $tc$ back to $sc$. | Automatically |
| S45 | **Tool-based**: The proposed smell detection rules in [86], [87] have been applied to each set of related metrics to detect smell instances for each smell. The authors have used the iPlasma tool [88] to generate eight required metrics. | No validation |

> **RQ3**: *What are the code smell dataset creation techniques and validation mechanisms?*
>
> **Summary for RQ3**: *Code smell datasets have been created with a manual or tool-based approach and validated manually with human experts, automatically with existing smell detection tools, or a combination of them. The number of studies that use tool-based approaches is increasing. Nearly 69% of the papers have used an automatic approach to create the code smell dataset of which 77% have leveraged only one tool, threatening the reliability of these datasets. The manual, automatic, and hybrid validations have respectively been used in 49, 9, and 2% of code smell datasets while 40% of the datasets have not been validated.*

#### 4.3.3  Software tools used to create code smell datasets

To answer the fourth research question, *which software tools are mostly leveraged to automatically create code smells datasets*, we extracted all tools mentioned in the primary studies. Various tools have been used to detect code smells and extract the source code metrics in preparing code smell datasets. We found 31 different software analysis tools vastly used in the code smell datasets creation process by researchers of the primary studies. Figure 3 shows the software tools used to detect the code smells and label smelly entities, *i.e.*, methods, classes, and packages by related smell types with the goal of dataset creation. DÉCOR [57] has been applied more than other available smell detection tools. The authors of two studies, S1 and S2, have developed their specific tools to label dataset samples. The tools used by each primary study are shown in Table 5. It should be noted that this table shows smell detection tools mentioned in the primary studies which used a tool-based method to create their datasets. Indeed, primary studies with manually created datasets have not been reported in Table 5. At most, five different tools have been simultaneously used by three primary studies, S4, S5, and S38.

Code smell datasets often contain additional information such as source code metrics for data samples. Figure 4 shows the reuse rate of the software tools used to compute code metrics corresponding to each sample in code smells datasets. Most tools have appeared only in one study. The POM[1] (Primitives, Operators, Metrics) tool [89] has the maximum reuse rate by appearing in three studies. POM is an extensible tool based on the PADL meta-model, which computes more than 60 metrics, including the well-known set of source code metrics by Chidamber and Kemerer [90]. The software tools that compute source code metrics may produce different results due to different definitions of metrics and calculation algorithms [38]. For example, authors in S14, have used two different metric calculation tools, CKJM [74] and POM [89], to calculate 62 source code metrics for each class. Some metrics have been calculated by both tools but the authors have decided to keep both versions since they are calculated differently in each tool, and thus, produce different values.

In addition, existing tools support a subset of source code metrics. For example, the CKJM tool [75] only calculates Chidamber and Kemerer [90] object-oriented metrics by processing the bytecode of compiled Java files to increase the accuracy and performance of metric computation. For this reason, applying multiple tools to compute source code metrics is recommended to increase the diversity and accuracy of metrics associated with each sample in code smell datasets. It is worth noting that most studies have not embedded source code metrics in their datasets. Source code metrics are used to detect code smells in different techniques, including machine learning, rule-based, and heuristic-based. Therefore, a decent code smell dataset is expected to be associated with various source code metrics.

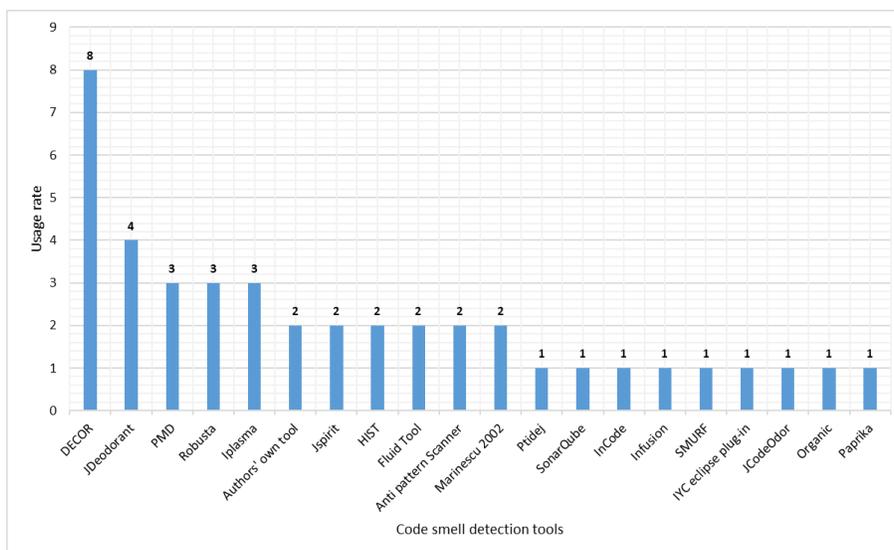

Figure 3. Tools used for creating code smell datasets.

---

[1]https://wiki.ptidej.net/doku.php?id=pom

**A systematic literature review on the code smells datasets and evaluation techniques**           *Zakeri-Nasrabadi et al.*

Table 5. Code smell detection tools used in each primary study.

| Tool | S1 | S2 | S3 | S4 | S5 | S6 | S8 | S10 | S11 | S12 | S15 | S17 | S19 | S21 | S22 | S23 | S24 | S25 | S28 | S29 | S30 | S31 | S32 | S33 | S34 | S35 | S36 | S37 | S38 | S39 | S40 | S42 |
|---|---|---|---|---|---|---|---|---|---|---|---|---|---|---|---|---|---|---|---|---|---|---|---|---|---|---|---|---|---|---|---|---|
| Authors' own tool | ✓ | ✓ | | | | | | | | | | | | | | | | | | | | | | | | | | | | | | |
| Manual | | | ✓ | | | ✓ | | ✓ | | | | ✓ | | | | ✓ | ✓ | | | ✓ | ✓ | | | | | | | | | | | |
| iPlasma | | | | ✓ | ✓ | | | | ✓ | | | | | | | | | | | | | | | | | | | | | | | |
| PMD | | | | ✓ | ✓ | | | | | | | | | | | | | | | | | | | | | | | ✓ | | | | |
| Fluid-Tool | | | | ✓ | ✓ | | | | | | | | | | | | | | | | | | | | | | | | | | | |
| Anti-pattern Scanner | | | | ✓ | ✓ | | | | | | | | | | | | | | | | | | | | | | | | | | | |
| Marinescu 2002 | | | | ✓ | ✓ | | | | | | | | | | | | | | | | | | | | | | | | | | | |
| Ptidej | | | | | | | ✓ | | | | | | | | | | | | | | | | | | | | | | | | | |
| SonarQube | | | | | | | ✓ | | | | | | | | | | | | | | | | | | | | | | | | | |
| DÉCOR | | | | | | | | | | ✓ | | | ✓ | | | | | | | | | | ✓ | ✓ | ✓ | | | | ✓ | ✓ | | ✓ |
| HIST | | | | | | | | | | | | | ✓ | | | | | | | | | | | | ✓ | | | | | | | |
| InCode | | | | | | | | | | | | | ✓ | | | | | | | | | | | | | | | | | | | |
| JDeodorant | | | | | | | | | | | | | ✓ | ✓ | | | | | | | | | | ✓ | | | | | ✓ | | | |
| Robusta | | | | | | | | | | | | | | ✓ | | | | | | | | | | ✓ | | | ✓ | | | | | |
| Infusion | | | | | | | | | | | ✓ | | | | | | | | | | | | | | | | | | | | | |
| SMURF | | | | | | | | | | | | | | | ✓ | | | | | | | | | | | | | | | | | |
| IYC eclipse plugin | | | | | | | | | | | | | | | | | | ✓ | | | | | | | | | | | | | | |
| Jspirit | | | | | | | | | | | | | | | | | | | | | | | | | | | ✓ | | ✓ | | | |
| JCodeOdor | | | | | | | | | | | | | | | | | | | | | | | | | | | ✓ | | | | | |
| Organic | | | | | | | | | | | | | | | | | | | | | | | | | | | ✓ | | | | | |
| Paprika | | | | | | | | | | | | | | | | | | | | | | | | | | | | | | | ✓ | |

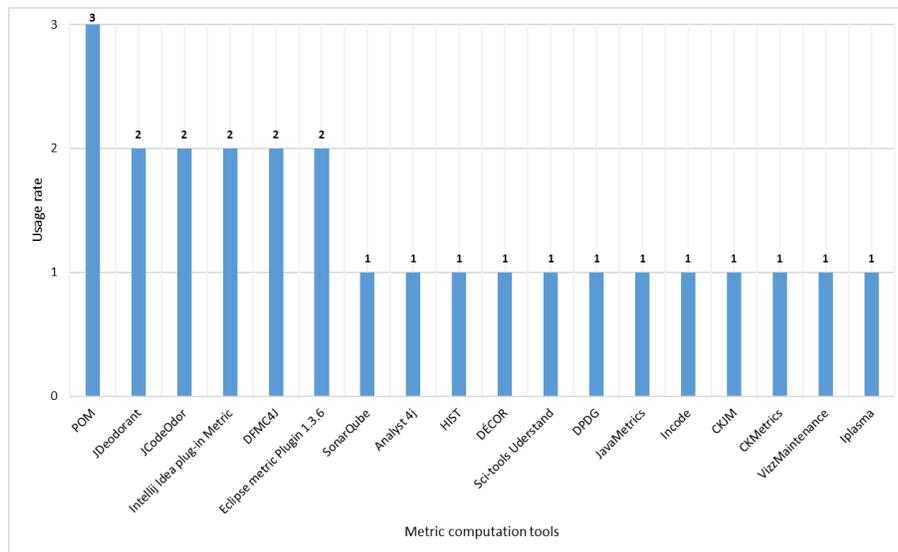

Figure 4. Tools used for extracting and computing source code metrics.

**RQ4**: *Which software tools are mostly leveraged to automatically create code smells datasets?*

**Summary for RQ4**: *At most, five different tools have been simultaneously used to create code smell datasets. DÉCOR [57], JDeodorant [81], and iPlasma [88] are the top three frequently used tools in creating code smell datasets. POM is mainly used to compute source code metrics required for smell detection. Most available code smell datasets do not have any precomputed source code metrics, making it time-consuming to create new smell detection tools based on them.*

### 4.4 Structural aspects of code smell datasets

This section answers RQ5 by analyzing the structural properties of the code smell dataset proposed in the primary studies. According to Figure 2 in Section 4.2, 43 out of 45 datasets are based on the Java language. However, their other structural properties such as the type of smells and the number of samples are different. We observed that existing code smell datasets do not follow a standard structure and often contain different metadata, making it difficult to fairly compare the datasets. The data and metadata are typically saved into XLS, CSV, SQL, or TXT files with required information about code smell, metrics, and projects. As an example, Figure 5 illustrates the structure and samples of the code smell dataset proposed in S10 [34]. The dataset contains 15 columns declaring various information about available samples. It contains nearly 15000 code samples of the smelly and non-smelly instances. Some code smell datasets contain source code metrics, which can be used as features by the code smell detection tools. If a dataset does not have any source code metrics, the link to the source codes of samples must be provided to extract the required features and metrics.

| id | reviewer_id | sample_id | smell | severity | review_timestamp | entity | code_name | repository | commit_hash | path | start_line | end_line | link | is_from_industry_relevant_project |
|---|---|---|---|---|---|---|---|---|---|---|---|---|---|---|
| 710 | 7 | 6236496 | long method | major | 17:30.9 | function | org.eclipse.elk.core.meta.ide.contentassist.antlr.internal.InternalMetaDataParser#rule__XWhileExpression_Group_0_Impl | git@github.com:eclipse/elk.git | 9a87764f00d863463b1be6de1920d8aa3c3ade70 | /plugins/org.eclipse.elk.core.meta.ui/src-gen/org/eclipse/elk/core/meta/ide/contentassist/antlr/internal/InternalMetaDataParser.java | 45156 | 45191 | https://github.com/eclipse/elk/blob/9a87764f00d863463b1be6de1920d8aa3c3ade70/plugins/org.eclipse.elk.core.meta.ui/src-gen/org/eclipse/elk/core/meta/ide/contentassist/antlr/internal/InternalMetaDataParser.java/#L45156-L45191 | 1 |
| 711 | 7 | 6236496 | feature envy | none | 17:30.9 | function | org.eclipse.elk.core.meta.ide.contentassist.antlr.internal.InternalMetaDataParser#rule__XWhileExpression_Group_0_Impl | git@github.com:eclipse/elk.git | 9a87764f00d863463b1be6de1920d8aa3c3ade70 | /plugins/org.eclipse.elk.core.meta.ui/src-gen/org/eclipse/elk/core/meta/ide/contentassist/antlr/internal/InternalMetaDataParser.java | 45156 | 45191 | https://github.com/eclipse/elk/blob/9a87764f00d863463b1be6de1920d8aa3c3ade70/plugins/org.eclipse.elk.core.meta.ui/src-gen/org/eclipse/elk/core/meta/ide/contentassist/antlr/internal/InternalMetaDataParser.java/#L45156-L45191 | 1 |
| 712 | 7 | 6234592 | long method | critical | 17:47.6 | function | org.eclipse.elk.core.meta.ide.contentassist.antlr.internal.InternalMetaDataParser#rule__MdCategory_Group_4_0_1_Impl | git@github.com:eclipse/elk.git | 9a87764f00d863463b1be6de1920d8aa3c3ade70 | /plugins/org.eclipse.elk.core.meta.ui/src-gen/org/eclipse/elk/core/meta/ide/contentassist/antlr/internal/InternalMetaDataParser.java | 22554 | 22599 | https://github.com/eclipse/elk/blob/9a87764f00d863463b1be6de1920d8aa3c3ade70/plugins/org.eclipse.elk.core.meta.ui/src-gen/org/eclipse/elk/core/meta/ide/contentassist/antlr/internal/InternalMetaDataParser.java/#L22554-L22599 | 1 |
| 713 | 7 | 6234592 | feature envy | none | 17:47.6 | function | org.eclipse.elk.core.meta.ide.contentassist.antlr.internal.InternalMetaDataParser#rule__MdCategory_Group_4_0_1_Impl | git@github.com:eclipse/elk.git | 9a87764f00d863463b1be6de1920d8aa3c3ade70 | /plugins/org.eclipse.elk.core.meta.ui/src-gen/org/eclipse/elk/core/meta/ide/contentassist/antlr/internal/InternalMetaDataParser.java | 22554 | 22599 | https://github.com/eclipse/elk/blob/9a87764f00d863463b1be6de1920d8aa3c3ade70/plugins/org.eclipse.elk.core.meta.ui/src-gen/org/eclipse/elk/core/meta/ide/contentassist/antlr/internal/InternalMetaDataParser.java/#L22554-L22599 | 1 |
| 714 | 8 | 6239751 | blob | minor | 18:33.6 | class | org.eclipse.elk.alg.mrtree.p2order.NodeOrderer | git@github.com:eclipse/elk.git | 9a87764f00d863463b1be6de1920d8aa3c3ade70 | /plugins/org.eclipse.elk.alg.mrtree/src/org/eclipse/elk/alg/mrtree/p2order/NodeOrderer.java | 37 | 169 | https://github.com/eclipse/elk/blob/9a87764f00d863463b1be6de1920d8aa3c3ade70/plugins/org.eclipse.elk.alg.mrtree/src/org/eclipse/elk/alg/mrtree/p2order/NodeOrderer.java/#L37-L169 | 1 |
| ... | ... | ... | ... | ... | ... | ... | ... | ... | ... | ... | ... | ... | ... | ... |

Figure 5. The structure (columns) and samples (rows) of the proposed dataset in S10.

#### 4.4.1 Supported code smells

One of the main characteristics of code smell datasets is the different types of smells and anti-patterns supported by the datasets. We counted the number and types of smells in the available datasets to answer RQ5 about the code smells covered by existing datasets. For the datasets that were not publicly available, we relied on the statistic provided in their corresponding primary studies.

Table 6 shows the smell types in the code smell datasets proposed by the primary studies. The proposed dataset in S42 contains 29 types of code smells, which is the highest among the primary studies. The datasets in S8 and S1 with 23 and 13 types of code smells are in second and third place, respectively. Figure 6 shows the available code smells and the numbers of supporting datasets. Smells with appearance frequency one has been shown as a separate category for better visualization. It is observed that God/ large class, long/ brain method, feature envy, data class, and spaghetti code are among the top five supported code smells. They form 43% of total smells covered in code smell datasets. One possible reason is the simplicity of manual detection and the number of available detection tools for these smells. On the other hand, a relatively large number of smells (nearly 16% of all smells listed in Table 6) are only supported by one dataset, demonstrating the lack of research on a large portion of code smells.

Table 6. Supported smell types by code smell datasets in primary studies.

| Study | Supporting code smells |
|---|---|
| S1 | Class data should be private, complex class, feature envy, God class, inappropriate intimacy, lazy class, long method, long parameter list, message chains, middle man, refused bequest, spaghetti code, speculative generality |
| S2 | Class data should be private, complex class, feature envy, blob, inappropriate intimacy, lazy class, long method, long parameter list, message chains, middle man, refused bequest, spaghetti code, speculative generality |
| S3 | Divergent change, shotgun surgery, parallel inheritance, blob, feature envy |
| S4 | God class, data class, feature envy, long method |
| S5 | God class, data class, feature envy, long method |
| S6 | Long method, data class, God class |
| S7 | God class, data class, feature envy, long method |
| S8 | Duplicated code, blob, class data should be private, cyclomatic complexity, down casting, excessive use of literals, feature envy, functional decomposition, God class, inappropriate intimacy, large class, lazy class/freeloader, orphan variable or constant, refused bequest, spaghetti code, speculative generality, Swiss army knife, tradition breaker, excessively long identifiers, excessively short identifiers, excessive return of data, long method, too many parameters/ long parameters list |
| S9 | Feature envy, long method |
| S10 | Blob, data class, feature envy, long method |
| S11 | Data class, God class, brain class, brain method |
| S12 | God class, long method, spaghetti code, complex class, class data should be private |
| S13 | Blob, data class, feature envy, long method |
| S14 | Anti-singleton, blob, class data should be private, complex class, large class, lazy class, long method, long parameter list, message chains, refused parent bequest, speculative generality, Swiss army knife |
| S15 | Blob, God class, data class, feature envy, schizophrenic class |
| S16 | Lazy class, feature envy, middle man, message chains, long method, long parameter lists, switch statement |
| S17 | Blob, data class, feature envy, long method |
| S18 | Blob, spaghetti code, functional decomposition |
| S19 | God class, feature envy |
| S20 | Large class, lazy class, data class, parallel inheritance hierarchies, God class, feature envy |

**A systematic literature review on the code smells datasets and evaluation techniques**　　　　　　*Zakeri-Nasrabadi et al.*

| Study | Supporting code smells |
|---|---|
| S21 | Feature envy, long method, God class, empty catch block, unprotected main, dummy handler, nested try statement, careless cleanup, over-logging, exception thrown from finally block |
| S22 | Blob, Swiss army knife, functional decomposition, spaghetti code |
| S23 | Blob, functional decomposition, spaghetti code |
| S24 | Long method |
| S25 | Long method, big class, feature envy, delegator, lazy class |
| S26 | Clone (duplicate code) |
| S27 | Clone (duplicate code) |
| S28 | Blob, functional decomposition, spaghetti code |
| S29 | God class |
| S30 | Blob |
| S31 | God class, long method, type checking, feature envy, empty catch block, careless cleanup, unprotected main, exception thrown in finally block, nested try statement, dummy handler, over logging |
| S32 | Anti-singleton, blob, class data should be private, complex class, large class, lazy class, long method, long parameter list, message chains, refused parent bequest, speculative generality, Swiss army knife. |
| S33 | Blob, functional decomposition, spaghetti code, Swiss army knife |
| S34 | Blob, complex class, spaghetti code, shotgun surgery |
| S35 | empty catch block, dummy handler, nested try statements, unprotected main, careless cleanup, exception thrown in the finally block |
| S36 | feature envy, dispersed coupling, refused parent bequest, God class |
| S37 | God class, data class, brain method, shotgun surgery, dispersed coupling, message chains |
| S38 | God class, long method, feature envy, refused parent bequest |
| S39 | Blob, long method, feature envy, spaghetti code, misplaced class |
| S40 | Internal getter/setter (IGS), member ignoring method (MIM), hash map usage (HMU) |
| S41 | Blob, divergent change, shotgun surgery, parallel inheritance, feature envy |
| S42 | Abstract class, child class, class global variable, class one method, complex class only, controller class, data class, few methods, field private, field public, function class, has children, large class, large class only, long method, long parameter list class, low cohesion only, many attributes, message chains class, method no parameter, multiple interfaces, no inheritance, no polymorphism, not abstract, not complex, one child class, parent class provides protected, rare overriding, two inheritance |
| S43 | Shotgun surgery, message chains |
| S44 | Feature envy, long method, large class, and misplaced class |
| S45 | Brain class and brain method |

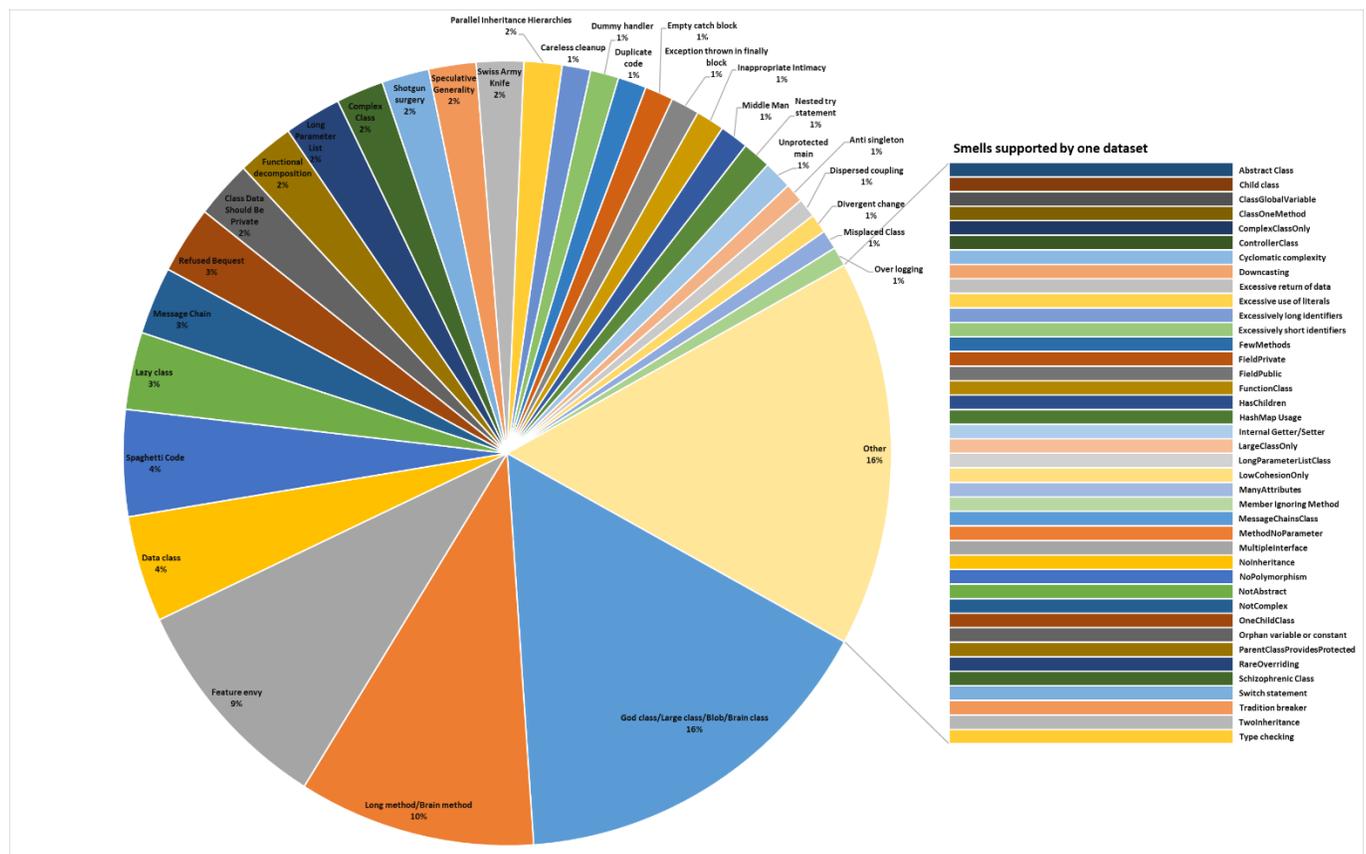

Figure 6. Frequency of smells and anti-patterns supported by available code smell datasets.

As shown in Figure 6, there are many types of code smells apart from those introduced by Fowler and Beck [1] and Lanza and Marinescu [86]. Moreover, some smells and anti-patterns have several alias names. We consider them as one type where possible. The lack of standard taxonomy for code smell types leads to various smell types proposed by datasets. Sometimes, the introduced type cannot be considered a code smell or is named improperly by the authors. For example, the cyclomatic complexity (CC) has been listed as a code smell in the dataset proposed by Lenarduzzi et al. [32]. However, CC is defined as a quality metric indicating the complexity of a program but not a code smell [91]. Creating code smell datasets based on accepted references such as [1], [86] is encouraged.

### 4.4.2 Instance ratio and features

Two other important structural aspects of code smell datasets are the number of instances and the number of metrics available in a dataset. Table 7 lists information about the size and metrics of the existing code smell dataset. A '—' symbol is used where the data are not available. Four datasets are balanced, *i.e.*, their number of smelly and non-smelly instances are equal while most existing datasets (40 datasets) do not have any non-smelly instances. Regarding metrics, 21 out of 45 datasets do not propose any additional metrics while the remaining dataset contains metrics at one or more entity levels including, method, class, package, and project. It is observed that there are no associated metrics for a large portion of code smell datasets. However, the code metrics can be extracted from the project's source code. For instance, the column named "link" in Figure 5 denotes the address of the source codes corresponding to the data sample in each row of the S10 dataset.

Figure 7 shows the instance ratio and diversity distribution of smell types in the existing code smell datasets. Only the five frequent smells in Figure 6 have been illustrated in this figure. Moreover, datasets without any reported data about samples' diversity have not been shown. The numbers in the vertical axis are reported in percentages for better comparison. It is observed that the ratio of smelly and non-smelly instances is not equal in most datasets. Similarly, the frequency of smell types is different in most datasets. For example, God class is more frequent than other types of smells. One possible conclusion is that such distribution mostly follows the natural distribution of code smells in software systems. Nevertheless, code smell datasets are expected to support various smell types regardless of their diversity.

Table 7. Code smell datasets instances and metrics.

| Study | Number of smelly and non-smelly samples | Number of code metrics |
|---|---|---|
| S1 | 17,350 smelly instances | 0 |
| S2 | 40,888 smelly instances | 0 |
| S3 | 243 smelly instances | 0 |
| S4 | • 4 dataset files, • 4×420 instances | • 61 metrics for class-level smells, • 82 metrics for method-level smells |
| S5 | • 4 dataset files • 4×420 instances | • 63 metrics for class-level smells, • 84 metrics for method-level smells |
| S6 | • 375 samples for each smell (125 smelly instances and 250 non-smelly instances) | 18 |
| S7 | • 4 datasets • 4×840 samples (140 instances and 700 non-smelly instances) | • 61 class-level metrics, • 82 method-level metrics |
| S8 | • 37,553 code smells, • 1,840,217 technical debt items | 30 |
| S9 | 445 samples | 46 metrics (20 class-level + 26 customized metrics) |
| S10 | • 14,739 (non-)smelly instances in total, • 974 blob instances, • 1,057 data class instances, • 806 long method instances, • 454 feature envy instances | 0 |
| S11 | 11,770 classes | 0 |
| S12 | 8,534 smelly instances | 9 |
| S13 | • 95 God class instances, • 77 Data class instances, • 114 feature envy instances, • 81 long method instances | — |
| S14 | 7,952 classes | 62 |
| S15 | • 700 method-level smells, • 190 class-level smells | — |
| S16 | 7 dataset files | 27 |
| S17 | • 600 samples, • 540 non-smelly instances, • 15 God class instances, • 15 data class instances, • 15 long method instances, • 15 feature envy instances | — |
| S18 | • 4381 non-smelly instances, • 431 smelly instances | — |
| S19 | • 73 God class instances, • 189 feature envy instances | • 4 metrics for God class, • 7 metrics for feature envy |
| S20 | 53,173 classes | 8 |
| S21 | — | — |
| S22 | 3,162 classes | 50 |
| S23 | • 777 classes, • 62 blob instances, • 52 spaghetti code instances, • 33 functional decomposition instances | — |
| S24 | 193 samples (37 long method instances and 156 non-smelly instances) | 4 |
| S25 | 688 classes | 9 |
| S26 | 281 code clone instances | — |
| S27 | 3,808 cloning operations | 21 |
| S28 | • 777 classes, • 19 Blob instances, • 19 functional decomposition instances, • 22 spaghetti code instances | — |
| S29 | 282 instances | — |
| S30 | • 777 classes (19 Blob instances and 758 non-smelly instances) | — |
| S31 | • 1,089 classes, • 429 God class instances, • 516 long method instances, • 131 type-checking instances, • 184 feature envy instances, • 73 empty catch block instances, • 36 careless cleanup instances, • 10 unprotected main instances, • 10 exceptions thrown in finally block instances, • 27 nested try statements, • 116 dummy handlers' instances | 17 |
| S32 | • 124,844 smelly instances • 69,345 non-smelly instances | 0 |

| Study | Number of smelly and non-smelly samples | Number of code metrics |
|---|---|---|
| S33 | • 998 blob instances, • 787 functional decomposition instances, • 779 spaghetti code instances, • 961 Swiss army knife instances | — |
| S34 | • 341 blob instances, • 349 complex class instances, • 313 spaghetti code instances, • 329 shotgun surgery instances | 20 |
| S35 | • 163 smell instances | — |
| S36 | — | 16 |
| S37 | • 590 smelly instances | 21 |
| S38 | • 14,918 smell instances, • 252,000 non-smell instances | • 17 class-level, • 13 method-level |
| S39 | • 4,267 smell instances | — |
| S40 | • 30 IGS, • 39 MIM, • 11 HMU | — |
| S41 | — | — |
| S42 | • 132,349 smelly instances • 36,217 non-smelly instances | 0 |
| S43 | • 459 shotgun surgery instances, • 483 Message chains instances, • 1,664 non-smelly instances | • 6 project-level, • 5 package-level, • 21 classes-level, • 19 methods-level |
| S44 | • 5049 misplaced class instances, • 12,976 long method instances, • 88,129 feature envy instances, • 86204 non-smelly instances | • 16 class-level, • 9 method-level |
| S45 | • 30 brain class instances, • 28 brain method instances | • 4 class-level, • 4 method-level |

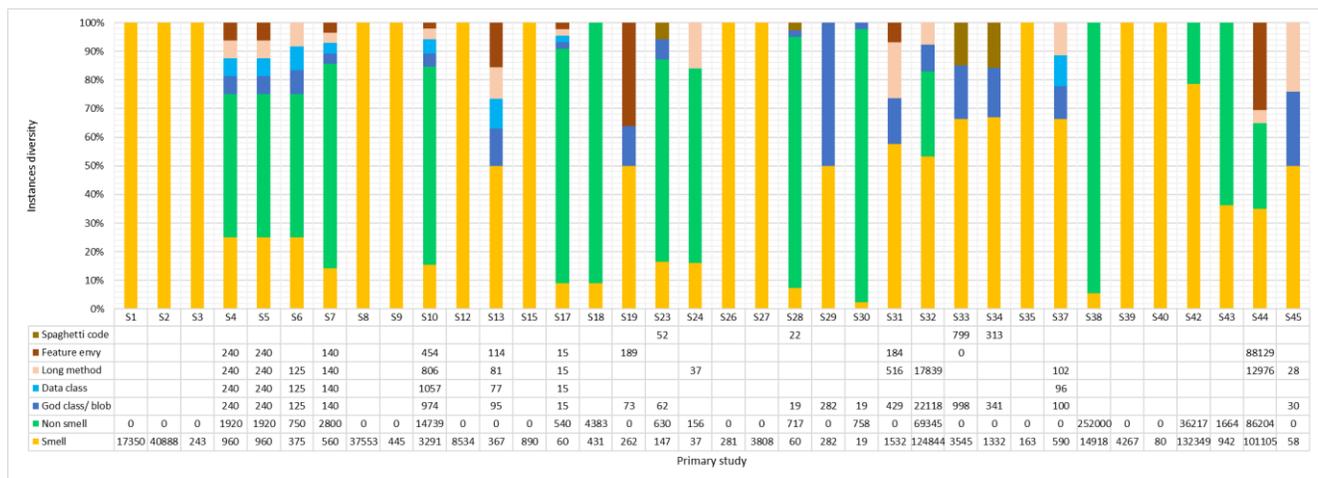

Figure 7. Instance ratio and distribution diversity of most frequent smells in existing code smell datasets.

**RQ5**: *Which programming languages, code smells, and code metrics are covered by existing datasets?*

**Summary for RQ5**: *We found that 43 out of 45 code smell datasets (i.e., 96%) are based on the Java programming language and contain smells related to object-oriented programs. At most, 70 types of smells are supported by available code smells datasets. God/ large class, long/ brain method, feature envy, data class, and spaghetti code are among the top five supported smells by the existing code smell datasets. Nearly 16% of all smells are only supported by one dataset, indicating the lack of oracles on a large portion of code smells. Four datasets are balanced while a majority of datasets (about 89%) do not contain any non-smelly instances. Moreover, 47% of code smell datasets do not contain any source code metrics.*

### 4.5 Source of data

The code smell datasets may build upon various software projects developed by academia, industry, and the open-source community. We extracted the names of all software projects mentioned in the primary studies to answer RQ6, *which open-source or close-source software projects are widely used as data sources to create code smells datasets?*

Figure 8 shows the frequency of project types and programming languages used in creating code smell datasets. Most researchers (nearly 89%) have prepared their datasets using open-source projects. However, few researchers (only 11%) have used industrial projects. We also did not find any set of academic projects explicitly developed for studying code smells in selected datasets. Industrial projects' participation in code smell datasets is expected to be increased in future research. Table 8 shows the list of projects used to create code smell datasets. As shown in Figure 9 word cloud illustration, the top five frequently used projects are Xerces[1] [92], Eclipse[2] [93], Gantt Project[3] [94], Argo UML[4]

---

[1] https://xerces.apache.org
[2] https://git.eclipse.org/c
[3] https://www.ganttproject.biz
[4] https://argouml-tigris-org.github.io/tigris/argouml

[95], Ant[1] [96], and JEdit[2] [97]. Moreover, the projects in the well-known Qualitas Corpus benchmark [98] have been used in five code smells datasets.

Most code smell datasets have been created based on the same set of software projects. One possible reason is that the datasets have the same authors who have preferred to work on the same software systems. The second is that the new dataset has only improved previous datasets by increasing the number of instances and adding some metrics or severity indexes to the dataset. Finally, the new datasets are created by merging the existing ones in some studies. Nevertheless, it is needed to consider new projects when creating new code smell datasets to increase the diversity of instances in different application domains.

Table 8. Source of data (benchmark projects) used to create code smell datasets.

| Study | Source projects | Types | Language |
|---|---|---|---|
| S1 | Apache ant, apache cassandra, apache derby, apache hadoop, apache hbase, apache hive, apache incubating, apache ivy, apache karaf, apache lucene, apache nutch, apache pig, apache qpid, apache struts, apache wicket, apache xerces, argo uml, atunes, eclipse core, elasticsearch, freemind, hibernate, hsqldb, jboss, jedit, jfreechart, jhotdraw, jsl, jvlt, sax | Open-source | Java |
| S2 | Apache ant, apache cassandra, apache derby, apache hadoop, apache hbase, apache hive, apache incubating, apache ivy, apache karaf, apache lucene, apache nutch, apache pig, apache qpid, apache struts, apache wicket, apache xerces, argo uml, atunes, eclipse core, elasticsearch, freemind, hibernate, hsqldb, jboss, jedit, jfreechart, jhotdraw, jsl, jvlt, sax | Open-source | Java |
| S3 | Apache ant, apache tomcat, jedit, android api (framework-opt-telephony), android api (frameworks-base), android api (frameworks-support), android api (sdk), android api (tool-based), apache commons lang, apache cassandra, apache commons codec, apache derby, eclipse core, apache james mime4j, google guava, aardvark, amd engine, apache commons io, apache commons logging, mongo db | Open-source | Java |
| S4 | 74 projects from qualitas corpus | Open-source | Java |
| S5 | 76 projects from qualitas corpus | Open-source | Java |
| S6 | 281 github projects | Open-source | Java |
| S7 | 74 projects from qualitas corpus | Open-source | Java |
| S8 | Acuumulo ambary, atlas, aurora, batik, beam, cocoon, commons bcel, commons beanutils, commons cli, commons codec, commons collections, commons configuration, commons daemon, commons dbcp, commons dbutils, commons digester, commons exec, commons file upload, commons io, commons jelly, commons jexl, commons jxpath, commons net, commons net, commons ognl, commons validator, commons vfs, felix, http components client, http components core, mina sshd, santuario java, zookeeper | Industrial | Java |
| S9 | 74 projects from qualitas corpus | Open-source | Java |
| S10 | 523 projects from GitHub | Industrial | Java |
| S11 | Tomcat, jurby, netty | Open-source | Java |
| S12 | Ant, argo uml, cassandra, derby, eclipse, elasticsearch, hadoop, hsqldb, incubating, nutch, qpid, wicket, xerces | Open-source | Java |
| S13 | Gantt project, xerces | Open-source | Java |
| S14 | Eclipse, mylyn, argo uml, rhino | Open-source | Java |
| S15 | Argo uml, jabref, jedit, mucommander | Open-source | Java |
| S16 | — | Industrial | Java |
| S17 | Gantt project | Open-source | Java |
| S18 | Argouml, xerces, ant-apache, azureus | Open-source | Java |
| S19 | Android opt telephony, android support, ant, lucene, tomcat, xerces, argo uml, jedit | Open-source | Java |
| S20 | Android-universal-image-loader, bigbluebutton, bukkit, clojure, dropwizard, elasticsearch, junit, libgdx, metrics, netty, nokogiri, okhttp, platform frameworks base, retrofit, presto, rxjava, spring-boot, spring, framework, storm, zxing | Open-source | Java |
| S21 | Mobac, jajuk, gogui, openrocket | Open-source | Java |
| S22 | Argo uml, azureus, xerces | Open-source | Java |
| S23 | Gantt project, xerces | Open-source | Java |
| S24 | Apache commons cli | Open-source | Java |
| S25 | Iyc, weka | Industrial | Java |
| S26 | Git, xz, bash, e2fsprogs | Open-source | C |
| S27 | Xproj, yproj (Microsoft projects) | Industrial | C# |
| S28 | Gantt project, xerces | Open-source | Java |
| S29 | Eclipse jdt, xerces | Open-source | Java |
| S30 | Gantt project, xerces | Open-source | Java |
| S31 | Jtds, jchess, artofillusion, ordrumbox | Open-source | Java |
| S32 | Eclipse, mylyn, argo uml, rhino | Open-source | Java |
| S33 | Argo uml, azureus, gantt project, log4j, lucene, nutch, pmd, quickuml, eclipse, xerces (two versions) | Open-source | Java |
| S34 | Apache mahout, apache Cassandra, apache Lucene, apache cayenne, apache pig, apache jackrabbit, apache jena, eclipse cdt, eclipse cfx | Open-source | Java |
| S35 | Apache abdera | Open-source | Java |
| S36 | Jhotdraw | Open-source | Java |
| S37 | Apache ant, apache camel, apache forrest, apache ivy, jedit, apache velocity, apache tomcat, apache lucene, apache pbeans, apache poi, apache synapse | Open-source | Java |
| S38 | 20 systems | Open-source | Java |
| S39 | Apache mahout, apache cassandra, apache lucene, apache cayenne, apache pig, apache jackrabbit, apache jena, eclipse cdt, eclipse cfx | Open-source | Java |

---

[1] https://ant.apache.org
[2] http://www.jedit.org

A systematic literature review on the code smells datasets and evaluation techniques  *Zakeri-Nasrabadi et al.*

| Study | Source projects | Types | Language |
|---|---|---|---|
| S40 | Soundwaves podcast, terminal emulator for android | Open-source | Java |
| S41 | Apache Ant, Apache Tomcat, jedit, and five projects belonging to the Android apis | Open-source | Java |
| S42 | Azureus and eclipse | Open-source | Java |
| S43 | 74 projects from the qualitas corpus | Open-source | Java |
| S44 | Junit, pmd, jexcelapi, areca, freeplane, jedit, weka, abdextractor, art of illusion, grinder | Open-source | Java |
| S45 | Argouml, aspectj, axion, batik, c-jdbc, cayenne, cobertura, collections, colt, columba, compiere, displaytag, drawswf, drjava, fitjava, fitlibraryforfitnesse, freecol, freecs, freemind, galleon, gantt project, hadoop, itext, james, javacc, maven, pmd, velocity, webmail, xmojo | Open-source | Java |

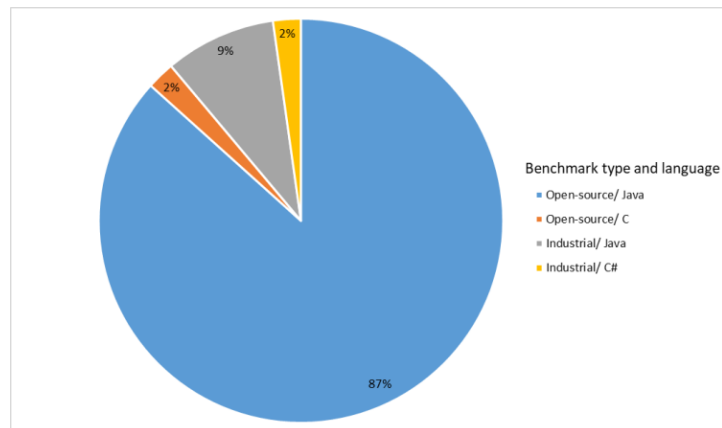

Figure 8. Benchmark project types and programming languages used in code smell datasets.

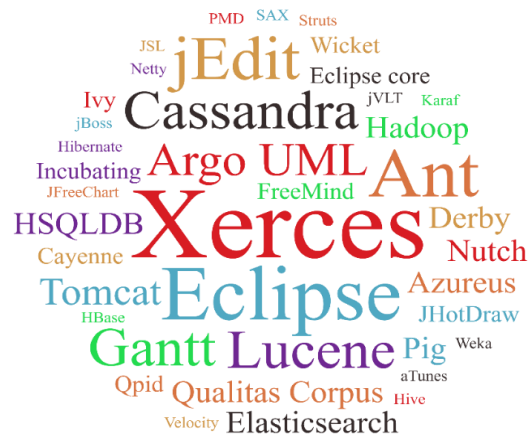

Figure 9. Word cloud of benchmark projects used to construct code smell datasets.

> **RQ6**: *Which open-source or close-source software projects are widely used as data sources to create code smells datasets?*
>
> **Summary for RQ6**: *About 89% of the code smells datasets have been created based on open-source projects, mainly including Xerces* [92], *Eclipse* [93], *Gantt Project* [94], *Argo UML* [95], *Ant* [96], *and JEdit* [97]. *These projects are also available in Qualitas Corpus* [98], *an old benchmark for code evolution studies. Existing code smell datasets suffer from the low diversity of projects, specifically projects in industrial domains.*

### 4.6 Code smell datasets' availability

Table 9 shows all publicly available code smell datasets and their download links. We have provided a complete reference for researchers and practitioners who aim to work on developing new code smell detection tools or datasets. It is observed that only 25 of 45 (56%) articles have proposed a public dataset and the remaining 20 datasets are not publicly available to download. The dataset links proposed in S3, S39, S40, and 41 were not accessible at the time of writing this SLR. It seems that these datasets are no longer supported by their authors or are no longer available for public usage. Hence, we marked the status of the S3, S39, S40, and S41 datasets as "obsolete" in Table 9. We conclude that the data used in a large portion of code smell detection research is not available, indicating that researchers are mostly not interested in publishing code smell datasets.

It is also observed that public code smell datasets have been published on different web repositories rather than well-known ones, such as Zenodo [99], Kaggle [100], and Figshare [101], which are specific to scientific datasets. It makes finding, indexing, updating, and versioning the datasets difficult. Currently, S7, S12, S34, and S37 are found in Figshare [101], S8, S9, S19, and S44 are found in GitHub [102], and S10, S20, and S44 are found in Zenodo [99]. Publishing datasets on the websites such as Zenodo [99], and Figshare [101], which provide data versioning facilities and statistics of datasets, helps researchers to find and select an appropriate dataset easily. A comprehensive review of the datasets in Table 9 facilitates using and extending these datasets. We discuss the most important contributions and properties of publicly available code smell datasets regarding their constructions and validation approaches in Section 5.

Table 9. List of publicly available code smell datasets.

| Study | Status | Projects | Supported smells | Other important properties | Download link |
|---|---|---|---|---|---|
| S1 | available | 30 | 14 | • 17,350 code smell instances, • .csv files | https://dibt.unimol.it/staff/fpalomba/reports/badSmell-analysis/index.html |
| S2 | available | 30 | 14 | • 40,888 code smell instances, • .csv files | http://www.mediafire.com/file/mzyr95cgmrbym19/dataset.zip/file |
| S3 | obsolete | 20 | 5 | 243 code smell instances | https://www.sesa.unisa.it/landfill/ |
| S4 | available | 74 | 6 | • 1,986 code smell and non-smell instances, • 61 metrics for class and 82 metrics for method, • .csv files | http://essere.disco.unimib.it/reverse/MLCSD.html |
| S5 | available | 76 | 4 | • 1,986 code smell and non-smell instances • 63 metrics for class smells and eight metrics for method smells, • 4 severity levels, • .csv files | http://essere.disco.unimib.it/reverse/MLCSD.html |
| S6 | available | 281 | 1 | • Long method instances, • PMD, Iplasma, Marinescu, Designite output for instance, • .csv files | http://madeyski.e-informatyka.pl/download/GrodzickaEtAl19DataSet.zip |
| S7 | available | 74 | 4 | • 4 datasets each 840 instances, • 140 with specific smell and 700 without it, • 61 class metrics and 82 method metrics, • .csv files | https://figshare.com/articles/dataset/Detecting_Code_Smells_using_Machine_Learning_Techniques_Are_We_There_Yet_/5786631 |
| S8 | available | 30 | 23 | • 37,553 code smell instances, • 1,840,217 technical debts, • 5 severity levels ,• 30 metrics, • .db file | https://github.com/clowee/The-Technical-Debt-Dataset |
| S9 | available | 74 | 2 | • 445 instances of multi-label smells and non-smells, • 2 labels, • 4 sets of labels, • 46 metrics, • .csv files | https://github.com/thiru578/Multilabel-Dataset |
| S10 | available | 523 | 4 | • 14,739 instances, • 3,291 code smell instances • 4 severity levels, • .xlsx files | https://zenodo.org/record/3666840#.YPbTXugzbIU |
| S12 | available | 13 | 5 | • 8,534 code smell instances and non-smell instances, • Naive Bayes and Décor outputs on instances, • .csv files | https://figshare.com/s/9d086c96eb1e58350f3d |
| S19 | available | 8 | 2 | • 262 smelly samples, • 11 metrics, • .txt files | https://github.com/antoineBarbez/SMAD |
| S22 | available | 3 | 4 | • 3,162 instances, • 50 metrics, •.csv and .arrf files | http://www.ptidej.net/download/experiments/wcre12a |
| S23 | available | 2 | 3 | • 777 instances, • 147 smell instances,• Décor output, • Metrics, • .xls files | http://www.ptidej.net/downloads/experiments/jss10 |
| S30 | available | 2 | 1 | • 777 instances, • 19 smell instances, • The probability that the class exhibiting the symptoms (smell) of a god class is truly a god class, • .xls files | http://www.ptidej.net/downloads/experiments/qsic09 |
| S32 | available | 4 | 12 | • Smell and non-smell instances, • Multi-label, • .csv files | http://www.ptidej.net/downloads/replications/emse10 |
| S33 | available | 11 | 4 | • Smell instances, • .ini files | http://www.ptidej.net/research/designsmells |
| S34 | available | 9 | 4 | • Smell instances, • Five levels of severity, • 20 metrics, • Including projects source code, • .csv files | https://figshare.com/s/94c699da52bb7b897074 |
| S37 | available | 11 | 6 | • Smell instances, • .csv files | https://figshare.com/articles/dataset/Toward_a_Smell-aware_Bug_Prediction_Model/4542709 |
| S38 | available | 20 | 4 | • Smell instances, • Result of multiple smell detection tools, • .csv files | https://dvscross.github.io/BadSmellsDetectionStudy/ |
| S39 | obsolete | 9 | 5 | • 4 severity levels, • 4,267 smell instances | https://goo.gl/GsPb1B,%E2%80%9D%202017 |
| S40 | obsolete | 2 | 3 | 60 smells instances | http://sofa.uqam.ca/paprika/mobilesoft16.phpCodeSmells |
| S41 | obsolete | 8 | 5 | — | http://www.rcost.unisannio.it/mdipenta/papers/ase2013/ |
| S42 | available | 2 | 29 | • Smells as features (multi-label), • .csv files, • High number of smells | https://www.ptidej.net/downloads/replications/wcre09a/ |
| S44 | available | 10 | 4 | • Refactoring opportunities binary labels, • .sql files | https://github.com/liuhuigmail/DeepSmellDetection |

**RQ7**: *What are the publicly available code smell datasets?*

**Summary for RQ7**: *Only 25 out of 45 (about 56%) of the code smell datasets are publicly available to be used by the research community. Therefore, most primary studies have created their own dataset, which is neither complete nor accurate. The dataset links of four primary studies are no longer publicly accessible. The available code smell datasets have been published on different websites and there is no dedicated host for publishing code smell datasets. As a result, code smell datasets maintenance, versioning, and metadata indexing are poorly supported by the community.*

## 4.7 Code smell datasets' quality, advantages, and disadvantages

This section discusses the answer to RQ8, *how is the quality of the existing code smell datasets regarding different evaluation metrics?* We observed that the quality of the dataset has been determined by different evaluation metrics, including accuracy, sensitivity, precision, and F1-score. The dataset validation mechanisms described in Section 4.3.2 aim at increasing code smell dataset quality. Theoretically, it is assumed that the labels of all samples in a dataset are true after validation is performed. Hence, the dataset is used as ground-truth to evaluate the code smell detection tools based on evaluation metrics. Indeed, the primary goal of the evaluation metrics is not used to evaluate the quality of the dataset, but actually, the performance of tools to identify code smells using the dataset. However, in practice, code smell datasets are not completely true due to a large number of instances, specifically the ones created automatically. In other words, the validation process most presumably does not remove all false positive and false negative instances. For this reason, evaluation metrics are used with a secondary goal of evaluating datasets, *i.e.*, reporting the number of false positives and false negatives in a dataset [53], [61].

Code smell analysis in software systems is often formulated as a binary classification problem [29], [60], [103] and evaluation metrics are computed according to the confusion matrix [104]. The confusion matrix [104] terminology used in primary studies for evaluating the results of code smell detection tools and code smell datasets quality is defined as follows:

- *True-positive (TP)*. TP refers to the entities, *e.g.*, methods or classes, that are smelly and considered smelly by a prepared dataset or tool.
- *True-negative (TN)*. TN refers to the entities that are not smelly and also not considered smelly by a prepared dataset or tool.
- *False-positive (FP)*. FP refers to the entities that are not smelly but considered smelly by a prepared dataset or tool.
- *False-negative (FN)*. FN refers to smelly entities which are not considered smelly by a prepared dataset or tool.

It should be noted that the discussed evaluation metrics are not specific to code smell detection with machine learning. We observed that primary studies reported their evaluation had used one or some of these metrics. Table 10 shows the evaluation metrics of the primary studies which at least reported one of the defined metrics. The F1 values are computed for all studies that have reported the Precision and Recall metrics. "NR" means that a specified metric has not been reported for that study. As discussed, the assumption about datasets is that they are ground-truth and fully accurate. Therefore, the quality metric values in primary studies have been mostly reported for tools considering such ground-truth datasets. Studies that reported evaluation metrics for their proposed tools, not their datasets, are marked with a "*" symbol. It is observed that 28 out of 45 studies reported at least one evaluation metric for their dataset or tools. Only two studies, S29 and S37, have reported the evaluation metrics for their dataset with an F1 score of 87 and 80%. The labels of the dataset created by using only one smell detection tool are as accurate as the tool. Therefore, we have reported the corresponding tool performance metrics for such datasets marked with a "+" symbol. Assessment of code smell datasets based on evaluation metrics indicates the presence of false labels in most of the datasets that are automatically labeled by code smell detection tools. Still, datasets created and validated manually can be considered highly precise compared to automatically created datasets.

Table 10. Evaluation metrics reported in primary studies.

| Study | Accuracy | Precision | Recall | F1 |
|---|---|---|---|---|
| S4* | 98.16 | NR | NR | 98.61 |
| S5* | 84.50 | NR | NR | NR |
| S6* | 97.79 | NR | NR | 98.33 |
| S7* | 76.00 | NR | NR | 10.00 |
| S9* | 97.50 | NR | NR | 97.60 |
| S12* | NR | 21.80 | 52.40 | 29.20 |
| S13* | NR | 80.30 | 76.20 | 78.20 |
| S14* | NR | 78.10 | 71.10 | 74.40 |
| S17* | 61.10 | NR | NR | NR |
| S18* | NR | 87.00 | 84.50 | 85.75 |
| S19* | NR | 42.50 | 66.50 | 51.90 |
| S20* | 99.09 | NR | NR | NR |
| S22* | NR | 83.27 | 80.57 | 81.90 |
| S26* | 70.00 | NR | NR | NR |
| S28* | NR | 73.24 | 100 | 84.55 |
| S29 | NR | 77.00 | 100 | 87.00 |
| S32+ | NR | 69.50 | 93.00 | 79.50 |
| S33+ | NR | 69.50 | 93.00 | 79.50 |
| S34* | NR | 82.25 | 81.25 | 81.50 |
| S36* | 91.38 | 92.04 | 90.26 | 91.37 |
| S37 | NR | 76.00 | 84.00 | 80.00 |
| S38* | 97.40 | 63.50 | 53.50 | 57.50 |
| S39+ | NR | 69.50 | 93.00 | 79.50 |
| S41* | NR | 76.20 | 76.40 | 76.20 |
| S42+ | NR | 69.50 | 93.00 | 79.50 |
| S43* | 99.80 | 99.80 | 99.80 | 99.80 |
| S44* | NR | 43.53 | 85.50 | 57.69 |
| S45* | 97.39 | NR | NR | NR |

The quality of the code smell datasets can be qualitatively evaluated regarding their advantages and disadvantages. Table 11 compares the advantages and disadvantages of existing code smell datasets. The main advantages and disadvantages of the manual oracle creation approaches are the high reliability and small size of the datasets, respectively, while for the automatic or tool-based methods, the opposite ones are true. The dataset validation is considered an advantage that supports its quality. Moreover, only seven out of 45 primary studies support the code smell severity in their proposed datasets making them superior to other datasets regarding this feature.

Table 11. Advantages and disadvantages of existing code smell datasets.

| Study | Advantages | Disadvantages |
|---|---|---|
| S1 | (1) Numerous samples, (2) Various smell types, (3) High recall | (1) No metrics, (2) Many duplicate instances, (3) Only smelly instances |
| S2 | (1) Numerous samples, (2) Various smell types, (3) High recall | (1) No metrics, (2) Many duplicate instances, (3) Only smelly instances |
| S3 | (1) Two experts | (1) No metrics, (2) Small dataset, (3) Only smelly instances |
| S4 | (1) Multiple tools and experts, (2) High number of metrics | (1) Old projects, (2) Balanced dataset, (3) One kind of smell in each dataset file |
| S5 | (1) Multiple tools and experts, (2) High number of metrics, (3) 4 levels of severity | (1) Old projects, (2) Balanced dataset, (3) One kind of smell in each dataset file |
| S6 | (1) Multiple experts | (1) Balanced dataset |
| S7 | (1) Imbalanced dataset, (2) Multiple types of smells, (3) High number of metrics | (1) Small size (few samples), (2) No validation |
| S8 | (1) Various smell types, (2) Numerous samples, (3) 5 levels of severity, (3) 30 different software metrics | (1) Only smell instances, (2) No validation |
| S9 | (1) Multilabel dataset | (1) Small size (few samples), (2) No validation |
| S10 | (1) Multiple expert developers, (2) 4 levels of severity | (1) No metrics |
| S11 | (1) Removing false positive instances automatically | (1) No metrics, (2) Small dataset, (3) Low recall (only one detection tool) |
| S12 | (1) High precision | (1) Balanced, (2) Low recall (only one detection tool has been used) |
| S13 | (1) Two metric computation tools | (1) No detailed information about the oracle creation |
| S14 | (1) Multilabel dataset, (2) Various smell types, (3) High number of metrics, (4) Two metric computation tools | (1) No validation |
| S15 | (1) Severity scores | (1) Low recall (Only one detection tool has been used), (2) Only smelly instances, (3) No validation |
| S16 | (1) Merging different datasets | (1) No information about smell detection or instance quantities, (3) No validation |
| S17 | (1) 40 expert developers, (2) High precision | Small size (few samples and smell types) |
| S18 | (1) "Artificial" code smell examples, (2) Reducing the manual effort effectively | (1) No information about smell detection, (2) No validation |
| S19 | (1) Using a weighted vote over the reported answers, (2) Using a couple of smell detection tools (high recall) | (1) Small dataset, (2) Only smelly instances |
| S20 | (1) High recall | (1) No validation |
| S21 | (1) Various types of smells | (1) No validation |
| S22 | (1) Interactive labeling | (1) No detailed information about the dataset |
| S23 | Seven experts (students) | (1) No metrics, (2) Small size (few samples and smell types) |
| S24 | (1) 5 levels of severity, (2) High precision | (1) One kind of smell, (2) Small size (few samples and smell types) |
| S25 | (1) Interactive labeling, (2) manual validation | (1) Small size (few samples and smell types) |
| S26 | (1) Interactive labeling | (1) One kind of smell, (2) Small size (few samples), (3) Only smell instances |
| S27 | (1) 21 different software metrics | (1) One kind of smell, (2) No validation, (3) Only smelly instances |
| S28 | (1) Merging existing datasets | (1) Small size (few samples) |
| S29 | (1) 4 experts (students) | (1) One kind of smell, (2) Small dataset, (3) Only smelly instances |
| S30 | (1) 4 experts (students) | (1) One kind of smell, (2) Small dataset |
| S31 | (1) Various smell types, (2) 21 different software metrics | (1) No validation |
| S32 | (1) Various smell types, (2) Multilabel | (1) No metrics, (2) No validation, (3) Low recall (only one detection tool) |
| S33 | (1) Well-defined process | (1) No metrics, (2) No validation, (3) Low recall (only one detection tool) |
| S34 | (1) 5 levels of severity, (2) Validated and ranked by the developers of the projects, (3) High reliability | — |
| S35 | (1) Different versions of the same classes | (1) No validation, (2) Small size (few samples) |
| S36 | — | (1) No validation, (2) No information about dataset size |
| S37 | (1) 5 levels of severity | (1) Small size (few samples) |
| S38 | (1) Lightweight and relatively reliable validation | — |
| S39 | (1) 4 levels of severity | (1) No validation |
| S40 | (1) Smells in the Android-based programs | (1) Small size (few samples) |
| S41 | (1) Two independent experts (students) | (1) No direct information about dataset size, (2) No metrics |
| S42 | (1) Smells as features (multi-label), (2) High number of smells | (1) No validation |
| S43 | (1) Automatic and relatively reliable validation | (1) No metrics |
| S44 | (1) Automatic and relatively reliable validation (2) Numerous samples | (1) No direct information about dataset size |
| S45 | — | (1) No validation, (2) No direct information about dataset size |

**RQ8**: *How is the quality of the existing code smell datasets regarding different evaluation metrics?*

**Summary for RQ8**: *The primary metrics used in evaluating the code smell datasets are accuracy, sensitivity, precision, and F1 score. An F1 score of 87 and 80% have been reported for datasets in S29 and S37 while no metrics have directly been provided for other datasets. The accuracy of datasets created by code smell detection tools is the same as the accuracy of the tool. We conclude that there is no fully accurate code smell dataset based on which code smell detection tools can be compared fairly. Moreover, no dataset is superior to other ones in all code smell dataset aspects.*

## 5 NOTABLE CODE SMELL DATASETS

This section answers RQ9, *what are the most comprehensive and adequate code smells datasets?* To this aim, we present an in-depth review of the most notable code smell detection datasets listed in Table 9. The discussed datasets in this section are either mostly cited by the author researchers in the field [29], [30], proposing a completely new labeling approach [13], [55], having a distinguished advantage [26], [27], [32], or addressing the problems of previous datasets [3], [34], [41]. Our review explains how code smells have been detected and to which extent the proposed datasets are valid. We end up with specific guidelines facilitating the creation and validation of code smell datasets mostly achieved by researchers in the field.

In the case of the manually created datasets, Fontana et al. [29] have developed a dataset containing 420 samples of four code smells from 76 Java projects in the Qualitas Corpus [98] by manual labeling. They asked a team of three M.Sc. students to identify the God class, long methods, feature envy, and data class smells in the selected projects and then label them with the corresponding smell type. Later they added a severity level for each kind of smell, including four ordinal levels [30]. The smelly and non-smelly samples have been balanced with a portion of 1/2 to be used in the machine learning task. However, studies show that realistic code smell datasets are highly imbalanced by nature due to the low occurrence of most code smells [4]. Therefore, Fontana's dataset seems unrealistic, and it also contains very few samples, such that it is not suitable to be used for learning-based smell detection techniques.

Di Nucci et al. [3] have criticized Fontana's dataset [29] concerning the size, types, and ratio of smelly and non-smelly samples. They created a dataset containing more than one type of smell and more samples by merging Fontana's datasets and reported that the performance of code smell detection models is up to 90% lower than the one reported in [29]. Their studies highlight the importance of code smell datasets concerning the number of samples and features, type of smells, and the ratio of smelly and non-smelly code samples. Moreover, it indicates that automatic smell detection is not a trivial task, and achieving high performance is difficult.

Madeyski and Lewowski [34] have stated that the projects in Qualitas Corpus [98] are old since they have been primarily developed with Java 5. Indeed, the features added in new versions of the Java programming language are not used in these projects. These features may lead to software smells that do not exist in the current datasets. They have introduced a dataset with 2,175 samples of 4 code smells, including God class, data class, feature envy, and long method, in 4 severity levels. The samples have been selected from industrial projects and labeled by 26 experienced software developers. Hozano et al. [41] have created a dataset containing 600 samples on four odor codes using 40 experienced software developers with at least three years of experience. Both datasets suffer from a low number of samples and smell types.

Some researchers have applied available tools and plugins to automatically create code smells datasets and expand different aspects of their datasets. Lenarduzzi et al. [32] have analyzed 33 Apache projects with the SonarQube[1] [73] and Ptidej [74] tools and extracted 23 types of smells in 5 severity levels. They have also extracted 30 source code metrics corresponding to each sample in the dataset by using SonarQube [73]. Tarwani et al. [55] have analyzed 1,089 Java classes in 4 projects with JDeodorant [81] and Robusta[2] [82] and recognized 11 smell types. Using an IntelliJ IDEA IDE plugin, they extracted a set of source code metrics corresponding to each sample. Using various tools in automated approaches leads to more code smells being detected and increased accuracy. However, these tools typically have a low agreement and produce many false positives.

Wang et al. [13] have proposed an approach to automatically detect code smells and remove false-positive samples. The authors have used RefDiff [105] to determine the refactored version of different entities (classes and methods) in code as a so-called contrastive version. The original version of the entities has been considered as the baseline. The authors have used the iPlasma tool [88] to identify code smells in both the original and contrastive versions. The smells found in the original version and refactored in the contrastive version have been labeled as smelly samples with their type. The entities that are not refactored and not detected as the smell have been considered non-smelly samples and added to the dataset. The smells detected by iPlasma [88] in the original version and not refactored in the contrastive version are false-positive and discarded. However, due to the low number of code smells, the approach proposed by Wang et al. may result in a high false-negative rate and aggravate data imbalance. To alleviate these problems, some researchers have employed a hybrid tool-based method in which false positives eliminates by human experts.

Palomba et al. [26] have created one of the largest code smell datasets using a tool-based approach in which smells are initially identified by a tool, and then false-positive samples are removed manually. Their smell detection tool uses a set of rules with strict thresholds to minimize the false positive rate. However, such a dataset may still suffer from a high false-negative rate. They have designated 17,350 samples of 13 code smell types from 395 versions of 30 different Java open-source projects. Unfortunately, the source code of 113 versions could not be found due to outdated links. Pecorelli et al. [4] have used this dataset to assess the role of data balancing in machine-learning-based code smell detection methods. In more recent work, Palomba et al. [27] have extended their previous dataset by increasing the number of samples to 40,888. Employing automatic approaches must not lead to the complete replacement of automatic methods with the human efforts required for creating reliable datasets. The results of automatic approaches must be deeply analyzed to discover how various code smells can be determined accurately. So far, the details proposed in many primary studies about creating and validating datasets are insufficient and must be revisited in future code smell studies, particularly those specific to smell datasets.

---

[1] https://www.sonarqube.org
[2] https://marketplace.eclipse.org/content/robusta-eclipse-plugin

Manually creating and validating code smell datasets is a laborious task. None of the smells are indeed trivial to find, and most of them have a subjective interpretation. The rules used to detect the smells are often proxies for them. For example, in the case of the most popular smell in the datasets, *i.e.*, God class, the detection strategy is based on the length of a source code file concerning all other source code files in the project. This strategy principally says nothing about clustering all functionality in one class. None of the papers really discuss the validity of this proxy. Other smells, such as the message chain or refused bequest, need far more complex analysis and understanding of the system to determine whether a snippet of code is affected by the smell. It means that any form of automation or manual analysis that relies on such proxies is almost always inaccurate. Using smell detection and dataset refinement in a cycle can help overcome such issues.

Experts in the study by Fontana et al. [29] reached a set of guidelines determining the most relevant aspects for each code smell to help the way labels are assigned. Lanza and Marinescu [86] observed that smelly codes often exhibit (*i*) low cohesion and high coupling, (*ii*) high complexity, and (*iii*) extensive access to the data of foreign classes. These observations are highly in common with the guidelines declared by Fontana et al. [29]. Table 12 summarizes guidelines used to recognize and validate the most frequent smells in code smell datasets. It concludes that a set of rules on which there is consensus can be extracted by analyzing smelly codes manually.

The code smell detection tool used by Palomba et al. [61] relies on detection strategies similar to those defined by Lanza and Marinescu [86]. Each detection strategy is a logical composition of predicates, and each predicate is based on an operator that compares a metric with a threshold. The detection strategy should be compared with those obtained by analyzing detection models, *e.g.*, the results of interpreting machine learning models, and then refined to achieve reliable smell detection guidelines for manual validation tasks. Finally, technical documentation of the various aspects of code smell datasets is necessary to provide useful guidelines when constructing new datasets.

Table 12. Guidelines to recognize and validate the most frequent smells in code smell datasets.

| Code smell/ anti-pattern | Detection guidelines |
|---|---|
| God/ large class, blob | • God/ large classes and blobs are large,<br>• God/ large classes expose a large number of methods,<br>• God/ large classes and blobs usually contain brain methods,<br>• God/ large classes and blobs tend to access many attributes from many other classes,<br>• God/ large classes and blobs tend to centralize the intelligence of the system. |
| Long/ brain method | • Long/ brain methods contain many lines of code,<br>• Long/ brain methods tend to have many parameters or a long parameter list,<br>• Long/ brain methods access many attributes and a large portion of the attributes declared in the enclosing class,<br>• The number of accessed variables through an accessor is high in long/ brain methods,<br>• Long/ brain methods tend to be complex. |
| Feature envy | • Feature envies access to many foreign attributes,<br>• Feature envies often access more foreign attributes than local ones,<br>• Feature envies mainly use the attributes that belong to a small number of other classes,<br>• Feature envies increase coupling and decrease cohesion. |
| Data class | • Data classes do not contain complex methods,<br>• Data classes expose few non-accessor methods and they are often very simple,<br>• Data classes mainly expose accessor methods,<br>• Data class attributes are often public or exposed through accessor methods. |

**RQ9**: *What are the most comprehensive and adequate code smells datasets?*

**Summary for RQ9**: *The proposed dataset by Palomba et al. [26] in S1 and the one proposed by Madeyski et al. [34] in S10, namely MLQC, can be considered the most comprehensive available code smell dataset according to the different aspects, mainly the size and quality of data samples. It is observed that publicly available code smell datasets have not been well-documented with respect to their structure, construction, and validation process.*

## 6 CHALLENGES, IMPLICATIONS, AND OPPORTUNITIES

Code smell datasets and detection tools face many challenges in reaching the completeness and reliability required by software engineers in the industry. In the previous sections, we discuss some of these challenges. This section investigates the answer to our last research question and focuses on the limitations of existing code smell datasets and the challenges of creating new ones. We describe six orthogonal dimensions of the most significant challenges and implications in current code smell datasets, which we observed while reviewing the primary studies. The possible solutions and opportunities are discussed to shed light on the direction of future research in this field. Each aspect of the code smell dataset, shown in Figure 2, can be improved by future works. This section ends up with our suggestion about an "ideal" dataset for code smells.

A systematic literature review on the code smells datasets and evaluation techniques      *Zakeri-Nasrabadi et al.*

### 6.1 Imbalanced smell types

Existing datasets support the limited and similar types of code smells. Figure 10 illustrates Fowler and Beck's catalog of code smells [1], covered by the current code smell datasets found in the primary studies along with their frequency. It is observed that code smells such as God/large class, long method, feature envy, and data class are supported by most datasets. However, six of the 22 code smells in Fowler and Beck's catalog are not supported by any datasets. These include data clumps, primitive obsession, alternative classes with different interfaces, incomplete library classes, temporary fields, and comments. Although Palomba's dataset [26] includes the comments code smell, they did not study this smell in their article.

The lack of datasets for the above smells and the ones proposed in Figure 6 in Section 4.4.1, which are only considered by one or two studies, prevents the development of comprehensive and reliable code smell detection tools. The available code smell detection or prediction tools only identify limited types of code smells with acceptable accuracy. It implies that only specific types of smells are expected to be found and fixed during the software maintenance phase.

It is important to note that not all smells require the same detection effort and difficulty. For instance, as we will discuss in Section 6.3, the long method is often easier to detect than the divergent change or message chain code smell. Indeed, the later smells in our example require more samples than the former, *i.e.*, the long method to be detected accurately. As a result, software developers have to perform manual investigations or use several time-consuming and error-prone tools to find code smells. The dataset by Khomh et al. [56] contains 29 code smell types, which is the highest number of smell types covered by the current code smell datasets. It implies that merging existing code smell datasets into one dataset while searching for samples of infrequent smells is necessary to improve the supported types of code smells in a dataset.

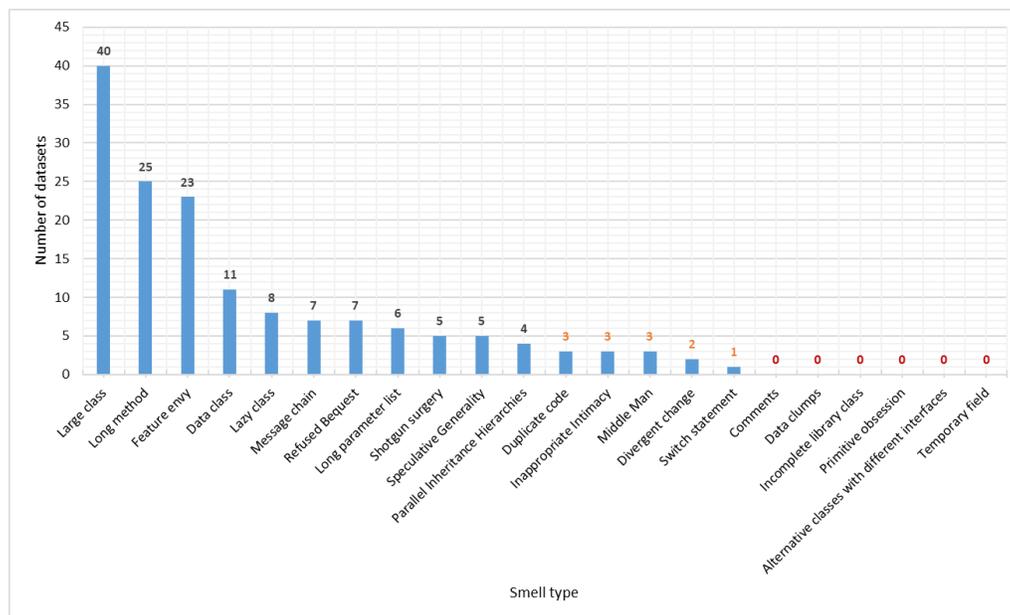

Figure 10. Code smells in Fowler and Beck's catalog [1] covered by the current code smell datasets.

### 6.2 Imbalanced smelly and non-smelly samples

It has been reported that the number of smelly samples is significantly less than non-smelly samples in real-world codebases [4]. The imbalanced data mainly affects learning-based approaches' performance in detecting code smells. Some researchers manually balanced smelly and non-smelly samples and built a balanced dataset to address this challenge [29], [30]. However, manually balancing leads to missing the approach's generalization and generates unrealistic results [3]. Pecorelli et al. [4] have examined different resampling algorithms, including ClassBalancer, Resample, SMOTE [106], and concluded that none of the resampling strategies could solve the imbalance problem in smell detection. It implies that synthetically generated samples only working at the feature level (*e.g.*, source code metrics) do not improve the effectiveness of learning-based smell prediction tools. A better solution to generating synthetically smelly samples that are more realistic than resampling feature vectors is to convert non-smelly code snippets into smelly ones by program transformation techniques. For instance, a long method can be created by merging several single responsible methods into one method using the inline method refactoring. The resultant feature vectors created by extracting the source code metrics from the refactored codes are more realistic than vectors generated by the resampling algorithms.

Another potential solution to address the problem of imbalanced data is to switch the learning paradigm from supervised learning to semi-supervised and unsupervised learning mechanisms. Specifically, the anomaly detection techniques are suitable if we define the problem of code smell detection as the problem of finding code snippets with anomalous or abnormal features. Such code snippets can be detected as outliers using an anomaly detection model such as isolation forest [107], local outlier factor (LOF) [108], or deep auto-encoder [109], [110].

In semi-supervised approaches, a large number of code snippets can be approximately labeled with their smells and then used in the final learning algorithm. The large-scale code smell datasets can be created with the described approaches and combined with the existing ones.

### 6.3 Different detection efforts and smell occurrences

Manual detection of the code smell is time-consuming and also requires high knowledge and experience in software development. Subjects with more professional backgrounds tend to reach higher precision regardless of their familiarity with the code they review [8]. The problem is exacerbated by the fact that different code smells required different detection efforts. For instance, detecting long method instances in a program can be performed by only looking at the method body. However, to detect the message chain code smell, the entire program call graph should investigate, requiring more effort than detecting the long method. In addition, some types of code smells rarely occur in the code, while they can have very harmful effects on software quality, and their detection is also a matter [111]. According to Tsantalis et al. [81] out of the 35,000 refactorings applied in the period 2011-2017, 50% belong to extract method, 25% correspond to extract class, and 16% correspond to move method. These refactoring operations respectively correspond to long method, God class, and feature envy which also frequently appeared in code smell datasets. The most frequent code smells in the Xerces-J v2.7.0 project fixed by Mkaouer et al. [112] are blob/God class, spaghetti code, feature envy, and data class which are the same reported in Figure 10 of our SLR. As a result, a positive correlation is observed between the frequency of supporting smells and their distribution in the projects. It should be noted that a low occurrence rate of a code smell should not lead to ignoring its impacts on software quality. Indeed, code smell datasets are expected to support such less common smells. For these reasons, the manually created datasets mainly suffer from both the low number of samples and the low number of supported smell types, specifically for those smells that rarely occurred.

Researchers have used automatic smell detection tools to address this problem. However, such automatically created datasets suffer from serious reliability issues, including high false positive and false negative rates, equal severities, limited smell types, and algorithmic biased. Moreover, dataset creation methods require a dynamic addition of new smell types effectively, which is not provided by current automatic approaches. Using multiple code smell detection tools, program version history, and code clone information reduces the false positive rate of automatically generated code smell datasets. Nevertheless, manual oracles are still required to archive reliable code smell detection tools, especially for new code smells. Studies show that even developers with little professional background can perform collaborative identification with high precision [8]. The available information made by developer actions in response to code smells, *e.g.*, refactoring and remodularization in the public code repositories, such as GitHub [113], speed up the manual labeling process. For example, comments on merging pull requests or the last comments on the issues which led to closing the issue often point out the change that applied to the code. The advanced NLP techniques can be used to detect those comments which denote the identification or refactoring of code smells and add the code with the corresponding smell to a dataset.

### 6.4 Different smell importance

Not all code smells are dangerous equally to the quality of the system [114]. For instance, the message chain negatively affects the testing and fault localization activities, while the data clumps do not have such destructive effects [111], [115]. This fact implies that it is necessary to create datasets containing information about the smells' impact on quality attributes. According to a recent survey by Lacerda et al. [116], limited empirical evidence about the impact of code smells on software quality attributes has been provided by the research community.

Determining the effects of code smells on different quality attributes such as reusability, understandability, and modularity requires both the frequency of code smells and the value of quality attributes. Computing accurate values for some quality attributes such as testability [117], [118] and Coverageability [119] is not straightforward since they require dynamic analysis which is a time and resource-consuming process. Adding quality attribute information to the code smell dataset facilitates the measurement of smell importance for different software systems.

The relationship between the code smells and quality attributes can be discovered by performing a correlation analysis or regression analysis. For instance, a regression model may be used to map a vector containing the frequency of different code smells to the value of the QMOOD quality attributes [120]. A feature importance analysis [121] is then applied to rank the smells by their importance in predicting the value of a specific quality attribute. A prerequisite of such meta-analysis is to focus on creating code smell datasets for the more critical types of smells. Code smells that are not covered by the existing smell datasets should be prioritized over the other smells, such as long method and God class, when creating new datasets. Future code smells datasets are expected to provide information about the importance of smell types on different aspects of code quality as golden references. This information enables the development of software tools that can identify the smells affecting specific tasks such as fault localization or fault prediction [122].

### 6.5 Different smell severity

Similar to the difference in the criticality of smell types, the instances of the same type also have different intensities [30], [123]. For example, two Java methods with 100 and 1000 lines of code are considered long methods [124], while the resultant technical debt imposed by them is very unlikely to be equal. Typically, smell detection tools use thresholds on different features, *e.g.*, related source code metrics, to select code smells that denotes a minimum value or lower bound for that feature. The distance between the specific value of a feature and the corresponding threshold for a given code snippet can be used to determine the smell severity.

Software engineers can effectively use the severity level of smells to prioritize the refactoring activities and reduce software maintenance costs and technical debts. Unfortunately, most available datasets do not provide the code smell severity levels or provide only two or three levels. As described in Table 11, only 7 out of 45 primary studies support the code smell severity in their proposed datasets. It implies that code smell datasets are rarely suitable for accurate estimation of technical debt and maintenance costs. The current code smells dataset should be improved to include information about the severity level of their smells to support smell prioritizations.

One solution to determine the severity of smells in the code smell dataset is to design and share online questionnaires with software developer communities and ask different developers to designate the severity of smell. In a long time, the approach leads to a reliable code smell dataset that supports the severity levels. Another solution is to find and analyze the number of refactored smells in the existing codebases during the software development lifecycle (SDLC). The refactored smells can be considered as smells with high priority from the developers' viewpoints and vice versa.

### 6.6  Diversity in application domains, programming languages, and paradigms

The last but not least point about the challenges in code smell datasets is related to three factors, including the application domain, programming language, and programming paradigm. Hall et al. [111] have stated that smells' impacts on systems depend on the application domain and development context. Our SLR reveals that the primary studies are limited to the concept of code smells and source code metrics in programming languages such as Java and C#, which are based on the object-oriented paradigm. There are vast opportunities to create code smell datasets for other programming languages, specifically multi-paradigm languages such as C++, Python, and Go, and even for the new generation of Java programming language [125] due to adding new features.

Although the current code smell datasets contain different software projects, they do not reflect any information regarding the application domain and development contexts. Fernandes et al. [126] have reported that the code smell detection tools present different levels of accuracy in different contexts. The critical software system, including business-critical, mission-critical, and safety-critical systems, rarely contributed to the existing code smell dataset. In contrast, the smell in such systems most presumably severely impacts system behavior and performance. It implies that the available code smells datasets are not dependable for critical systems.

Therefore, the next generation of code smell datasets is supposed to consider factors related to the application domain and development context, mainly the system criticalness level [127], deployment and run-time environments, and stakeholders. For instance, when voting between the result of tools, the application domain can be used to weigh each tool according to the application [126].

To figure out an "ideal" dataset for code smells, we back to our proposed classification in Section 4.2. An ideal dataset contains labels entirely validated, supports multiple programming languages, and cover various type of smells with a large number of samples, and features. The ratio of samples in the dataset must approach their ratio in reality. The minimum required features are smell importance, severity levels, sample sequences (to trace smell cooccurrence), application domain, and source code metrics. Additional metadata such as traceability links (to the source code of each sample), reviews activities, and versions are helpful. The dataset is expected to regularly update under a data version control system which enables tracking of different versions and covers the true samples of the previous datasets/ versions. Finally, the source of data, *i.e.*, the benchmark projects must be diverse as possible.

> **RQ10:** *What are the limitations of the existing code smell datasets?*
>
> **Summary for RQ10:** *Existing code smell datasets contain limited types of smells with few smelly samples and small sizes. In practice, the number of smelly samples is fewer than non-smelly ones, which leads to imbalanced datasets. Future research must develop new code smell datasets for supporting the severity and importance of smells, various application domains, programming languages, and paradigms. Using online questionnaires and generating synthetic smelly instances by program transformation are recommended to assist the code smells datasets creation process.*

## 7  THREATS TO VALIDITY

Several issues may threaten the construction, internal, and external validity of this paper. The main construction validity threat is the suitability of research questions and the categorization scheme used for answering these questions. We mainly focused on questions that researchers and practitioners face when developing and evaluating a new smell detection tool or comparing the existing ones to mitigate threats. The taxonomies used to classify the common aspects of code smell datasets were extracted from the taxonomies that appeared in the primary studies. A few factors, such as quantitative results of code smell detection obtained from each dataset, may still be missed in our research.

The internal validity of our paper may threat due to the incomplete set of articles selected as primary studies. Search engines, search terms, and inclusion/exclusion criteria are carefully defined to ensure that our review is comprehensive and the result is repeatable. Another problem we faced during the article selection process was missing some relevant papers that we expected to find in our initial search. Indeed, when feeding our search string to the search engines, especially the IEEE Xplore search engine, we noticed that some important papers containing our keywords were not found despite being indexed in their libraries. For example, our search string contains some of the keywords in the paper describing a smell detection tool, DÉCOR [57]. However, when feeding our search string to IEEE Xplore we noticed that it could not find the paper. We observed that similar problems have been reported by Landman *et al.* [128] when dealing with the IEEE Xplore search engine. Therefore, the problem was not specific to our search string. Fortunately, we could find such missed papers during the snowballing process.

We did not perform a strict quality assessment on the evaluation results of the primary studies due to the low number of relevant publications to mitigate the risk of losing any code smell dataset. However, we filtered the papers, which did not include introducing any new code smell dataset. In addition to digital search libraries, we look for code smell datasets in specific web repositories, including Zenodo [99], Kaggle [100], Figshare [101], and GitHub [113], to mitigate the threats of incomplete search.

Another threat to the internal validity is the manual analysis performed on the primary studies to extract the required information for analyzing and comparing existing code smell datasets. The initial analysis was performed independently by the first and third authors who are experienced in the software refactoring field. Tiny Python scripts were developed to help analyze public datasets by extracting primary statistics from the available versions. Afterward, objective information extracted from the primary studies was reviewed by three independent M.Sc. students in software engineering with a background in software smells and refactoring, and their correction was applied accordingly. Finally, the second and fourth authors also reviewed the results to ensure the collected result's correctness. We maintained the extracted information from the primary study in an Excel worksheet publicly available [25] to facilitate the analysis process and merging of results.

The external validity of our SLR is threatened by factors that affect the generality of reported results. We mainly relied on the statistics and results mentioned by the primary studies. Few code smell datasets appeared in more than one paper. We referred to them only if a selected primary study had not reported the required data. In some cases, the required data, *e.g.*, the size of the dataset or instance ratio, were not reported in any paper, and we could not find any relevant data. Indeed, ours is a work a meta-analysis of the datasets, we do not aim at producing additional knowledge other than shared by the original authors. Some code smells and anti-patterns have different names and slightly different definitions regarding the terminology [9]. We merged them into one type of smell based on the catalog of Fowler and Beck [1] and the recent SLR by Sharma and Spinellis [9] to mitigate the sparsity of the studied smell types.

## 8  CONCLUSION

High-quality and large code smells datasets are required to construct decent automatic smell detection tools and evaluate them. It is difficult to find and employ an appropriate code smell dataset to create new smell detection tools or evaluate the existing ones. A systematic literature review of the existing code smells datasets is proposed in this paper to answer ten research questions. Our research questions cover a wide range of information about code smell datasets, including their structure and formats, size, supporting languages, supporting smells, analysis tools, labeling approach, quality, and limitations. Code smells datasets are classified and studied in a standard model based on five different aspects in alignment with the research questions.

A total of 45 code smell datasets are recognized and reviewed, of which only 25 datasets are available publicly. Code smells datasets are created and evaluated manually, automatically, and semi-automatically. Existing code smells datasets suffer from limitations in the number of samples, supported smell types, and diversity of projects. Most datasets cover God class, long method, feature envy, and data class smells. At the same time, there is not any dataset for six of the smells discussed by Fowler and Beck [1]. The main reason is that the frequency of code smells in real-world projects and the effort required to find them are different. We observe that 43 out of 45 primary studies contain smelly instances for the Java programming language, indicating the lack of datasets for other programming languages and paradigms.

Reliable and large code smell datasets are mandatory artifacts in developing code smell detection and program refactoring tools, specifically the programs designed in the Software 2.0 paradigm [129], including learning-based software systems. With current datasets creating accurate learning-based tools are almost impossible. There are several opportunities for future work on code smell datasets. Generating and integrating code smell datasets automatically by composing different tools or synthesizing smelly samples by program transformation techniques [130] (*e.g.*, inverse refactoring [131]) can be considered new solutions and research lines in this area. One straightforward option to begin is to merge the current datasets into a single database to increase the size and types of smelly samples. Another is to leverage techniques such as code similarity detection and transfer learning to make datasets for other programming languages with a relatively low effort. Providing code smell datasets for new programming languages encourage researchers and practitioners to propose new and accurate smell detection tools with reliable evaluations.

## COMPLIANCE WITH ETHICAL STANDARDS

This study has received no funding from any organization.

## CONFLICT OF INTEREST

All of the authors declare that they have no conflict of interest.

## ETHICAL APPROVAL

This article does not contain any studies with human participants or animals performed by any of the authors.